\documentclass[fleqn,usenatbib]{mnras}

\usepackage{newtxtext,newtxmath}

\usepackage[T1]{fontenc}

\DeclareRobustCommand{\VAN}[3]{#2}
\let\VANthebibliography\thebibliography
\def\thebibliography{\DeclareRobustCommand{\VAN}[3]{##3}\VANthebibliography}

\usepackage{graphicx}	
\usepackage{amsmath}	
\usepackage{ulem}       
\graphicspath{{Pictures/}}

\newcommand{\de}{{\rm d}}
\newcommand{\msun}{{\rm M_\odot}}

\def\gsim{\rlap{\lower 2.5pt
 \hbox{$\sim$}}\raise 1.5pt\hbox{$>$}\;}
\def\lsim{\rlap{\lower 2.5pt
   \hbox{$\sim$}}\raise 1.5pt\hbox{$<$}\;}

\title[Massive seed black holes in quasar hosts]{Forming massive seed black holes in high-redshift quasar host progenitors}

\author[A. Lupi et al.]{
Alessandro Lupi,$^{1}$\thanks{E-mail: alessandro.lupi@unimib.it}
Zolt\'an Haiman,$^{2}$
and Marta Volonteri$^{3}$
\\
$^{1}$Dipartimento di Fisica ``G. Occhialini'', Universit\`a degli Studi di Milano-Bicocca, Piazza della Scienza 3, I-20126 Milano, Italy\\
$^{2}$Department of Astronomy, Columbia University, 550 West 120th Street, New York, NY, 10027, U.S.A.\\
$^{3}$Sorbonne Universit\`{e}s, UPMC Univ Paris 6 et CNRS, UMR 7095, Institut d'Astrophysique de Paris, 98 bis bd Arago, F-75014 Paris, France\\
}

\date{Accepted XXX. Received YYY; in original form ZZZ}

\pubyear{2021}

\begin{document}
\label{firstpage}
\pagerange{\pageref{firstpage}--\pageref{lastpage}}
\maketitle

\begin{abstract}
The presence of massive black holes (BHs) with masses of order $10^9\msun$, powering bright quasars when the Universe was less than 1 Gyr old, poses strong constraints on their formation mechanism. Several scenarios have been proposed to date to explain massive BH formation, from the low-mass seed BH remnants of the first generation of stars to the massive seed BHs resulting from the rapid collapse of massive gas clouds.
However, the plausibility of some of these scenarios to occur within the progenitors of high-z quasars has not yet been thoroughly explored.
In this work, we investigate, by combining dark-matter only N-body simulations with a semi-analytic framework, whether the conditions for the formation of massive seed BHs from synchronised atomic-cooling halo pairs and/or dynamically-heated mini-haloes are fulfilled in the overdense regions where the progenitors of a typical high-redshift quasar host form and evolve. Our analysis shows that the peculiar conditions in such regions, i.e. strong halo clustering and high star formation rates, are crucial to produce a non-negligible number of massive seed BH host candidates: 
we find $\approx1400$ dynamically heated metal-free mini-haloes, including one of these which evolves to a synchronised pair and ends up in the massive quasar-host halo by $z=6$.  This demonstrates that the progenitors of high-redshift quasar host haloes can harbour early massive seed BHs. Our results further suggest that multiple massive seed BHs may form in or near the quasar host's progenitors, potentially merging at lower redshifts and yielding gravitational wave events.

\end{abstract}




\section{Introduction}
The observations of quasars at redshift $z>6$ with massive black holes (MBHs) in excess of a few $10^8\msun$ \citep{fan06,mortlock11,banados18,decarli18} pose tight constraints on the formation and growth of these objects, and represent a challenge for theoretical models \citep[see, e.g.][for a recent review]{inayoshi20}.
Models of the evolution of the quasar luminosity function over the range $0<z\,\lsim 6$ suggest that the {\it seeds} of MBHs powering these quasars were in place early on, and subsequently grew via accretion and mergers
\citep[e.g.][]{soltan82,small92,kauffmann00,shankar09,shen20}.
To date, several scenarios have been proposed to explain the early formation of MBHs, either relying on established physical processes or in some cases based on unconventional physics. 
In the so-called `light-seed' scenario, BHs formed already at $z\gtrsim 20$ as remnants of the first generation of stars \citep[e.g.][Population III, hereafter PopIII]{haiman01,madau01a,heger03,volonteri03}. Relative to metal-enriched stellar populations, the initial mass function (IMF) of PopIII stars is still poorly constrained, due to i) the difficulty to observe this population in the local Universe \citep{schlaufman2018}, and ii) the uncertainties about the interplay between gas accretion/fragmentation and radiative feedback from the newly formed stars. 
 
Early results suggested that the first generation of stars were unusually massive, collapsing into BHs with masses up to a few 100~$\msun$ \citep{abel02,mckee08}. More recent studies found that fragmentation in the stellar accretion disc and radiative feedback reduce the typical masses of these stars, but with their IMF still top-heavy, producing remnant seed BHs in the range 10--1000$ \msun$ \citep{stacy12,hirano15,hirano17,hirano18,wollenberg20,kimura20}. 
Although these seeds can form at very early times, sustained accretion at the Eddington limit\footnote{The Eddington limit here refers to the accretion rate $\dot{M}_{\rm Edd}\equiv L_{\rm Edd}/c^2$ where $L_{\rm Edd}$ is the Eddington luminosity and $c$ the speed of light.} 
is required to reach the MBH masses observed in high-redshift quasars.  This condition is unlikely to be realistic, especially at early times, when galaxies are small and supernova feedback and/or stellar UV radiation can easily expel gas from the halo hosting the BH. 

A possible solution to this limitation is the occurrence of relatively short super-Eddington phases. Exceeding the Eddington rate even by a relatively modest factor, as expected in so-called slim disk models,  could allow initially small BHs to efficiently grow by orders of magnitude in mass in a few Myr~\citep{madau14,volonteri15,lupi16,pezzulli16}.  Given a sufficiently large inflow rate from larger radii, the Eddington rate can be exceeded by a much larger factor (several orders of magnitude), due to the trapping of the radiation in the inflowing plasma~\citep{begelman79}, resulting in rapid growth via "hyper-Eddington" accretion~\citep{volonteri05,tanaka09,pacucci15,inayoshi16,sakurai16}.  Caveats to this scenario include mechanical feedback from a jet,
which can significantly affect the dynamics of the accreting gas \citep{regan19}, and the need for the accreted gas to efficiently shed angular momentum (which may be facilitated in the early stages of growth by the presence of a nuclear star cluster; \citealt{alexander14}).

Another possible solution is the formation of intermediate-mass BHs with masses of about $10^4\msun$ in mildly metal-enriched nuclear stellar clusters, either via collisions between stars \citep{portegies02,gurkan04,omukai08,devecchi09,devecchi10,devecchi12,katz15,boekholt18,reinoso18,das20,tagawa20} or the runaway merger of stellar-mass BHs \citep{davies11,miller12,lupi14,tagawa15,kroupa20}. 
This range of masses is also obtained in models invoking physics beyond the standard model, for instance from the collapse of self-interacting~\citep{balberg02,pollack15} or other forms of dissipative dark matter \citep{damico18,latif19}. However, because of the relatively low initial mass of these seeds, it is not clear yet whether they can represent the seeds of the MBHs in high-redshift quasars. 

Finally, in the `heavy-seed' scenario, even more massive seeds can form, 
via the rapid 
collapse of a massive gas cloud \citep{oh02,koushiappas04,begelman06,lodato06,mayer10}.  Such a collapse may be facilitated by the presence of strong ultraviolet radiation that is able to suppress H$_2$ formation and cooling \citep{omukai01a,dijkstra08,agarwal12,latif13,visbal14b,latif15}, and likely proceeds through the intermediate stage of a supermassive star~\citep{hosokawa12,hosokawa13,woods17,haemmerle18}, or a quasi-star \citep{begelman08,begelman10,dotan11,choi13,fiacconi17b}. In these models, which have been collectively dubbed to produce a "direct collapse black hole (DCBH)", a strong UV radiation field is thought to be required to efficiently suppress H$_2$ formation in the haloes, the main cooling channel in cold pristine gas, able to trigger fragmentation and inhibit the monolithic collapse of $10^4-10^5\,\msun$. 

The large required UV intensity ($J_{21}\gsim 1000$ in units of 
$10^{-21}\rm\, erg\, s^{-1}\, cm^{-2}\, Hz^{-1}$; e.g.  \citealt{wolcott-green19}) is typically realised in the case of synchronised pairs, i.e. pairs of atomic cooling haloes (ACHs; separated by $\lsim 1$kpc) where one of the haloes forms stars a few Myr earlier which then illuminate the other one, inhibiting H$_2$ formation \citep{dijkstra08,agarwal12,dijkstra14,latif14,visbal14b,chon16,regan17,chon18}.

Another requirement in this scenario is that the gas avoids cooling and star-formation prior to reaching the ACH stage. H$_2$-free ACHs may be produced by a combination of a lower UV background flux ($J_{21}\sim 0.01-1$) and the unusually rapid assembly of a subset of ACHs: the corresponding reduction in ${\rm H_2}$ cooling \citep{fernandez14} and the accompanying increase in dynamical heating \citep{yoshida03,inayoshi15,inayoshi18} of gas in the mini-haloes (haloes with virial temperatures below the atomic cooling threshold, i.e. $T_{\rm vir}\lesssim 10^4$~K) could naturally result in star-free ACHs.  While these halos would form ${\rm H_2}$ and cool once they reach the ACH threshold, it has recently been suggested that they may nevertheless have large central gas inflow rates and form relatively massive seed BHs~\citep{wise19,sakurai20} of at least $10^3 \msun$ \citep{regan20b}.

In this work, we investigate the plausibility of the "synchronised-pair" and "dynamical-heating" scenarios (both separately and in combination) for the formation of massive seed BHs as progenitors of $z>6$ quasars.  While we build on previous semi-analytic studies by \citet{dijkstra14}, \citet{visbal14b}, \citet{habouzit16} and \citet{2016MNRAS.463..529H}, the novel feature of the present study is that we focus specifically on an overdense region of the Universe, which evolves into a massive ($\sim 10^{12}~{\rm M_\odot}$) dark matter halo by $z\approx6$, i.e. typical of haloes hosting bright quasars at this redshift.  This is achieved by combining semi-analytic prescriptions with N-body simulations resolving the merger history of such a massive halo.

The rest of this paper is organised as follows: in \S~\ref{sec:method} we summarise the underlying N-body simulations and present our improved semi-analytic framework, in \S~\ref{sec:results} we present our main results, in \S~\ref{sec:caveats} we discuss the caveats of our work, and in \S~\ref{sec:discussion} and \S~\ref{sec:conclusions} we discuss and summarise our conclusions and the implications of this work.

\section{Method}
\label{sec:method}


In this work, we investigate whether the so-called "direct-collapse" scenario can explain the formation of MBHs in high-redshift quasars. To this aim, we build a semi-analytic model on top of the high-resolution dark matter (DM) only equivalent of the hydrodynamic simulation presented in \citet{lupi19b}, a zoom-in simulation  targeting a high-sigma peak, i.e. a massive halo with $M_{\rm vir}\sim 3\times 10^{12}\,\msun$ at $z=6$. The initial conditions of our simulation consist of a high-resolution Lagrangian region of $\sim$6.5~cMpc per side at $z=100$ within a 100~Mpc comoving box, which shrinks down to 2.5~$R_{\rm vir}$ of the halo at $z=6$. For the current study, we stored outputs every $\Delta t=1$~Myr, which allows us to accurately check whether the different criteria for the direct collapse scenario are fulfilled.
The simulation was performed with \textsc{gizmo} \citep{hopkins15}, descendant of \textsc{gadget2} \citep{springel05}, assuming a $\Lambda$CDM Universe with the \citet{planck16} cosmology parameters, i.e. $\Omega_{\rm m}=0.3089$, $\Omega_\Lambda = 0.6811$, $\Omega_{\rm b}=0.0486$, $H_0=67.74\rm\, km\, s^{-1}\, Mpc^{-1}$, $\sigma_8=0.8159$, and $n_{\rm s}=0.9667$. The mass resolution is $m_{\rm DM}\approx 9\times 10^4\,\msun$, and the spatial resolution (Plummer-equivalent gravitational softening) is initially set to 640~pc comoving, which we recall corresponds to 640~pc/(1+z) in proper units, so that the resolution decreases with redshift, and switches to a constant resolution of 40~proper~pc below $z=15$.

For the current analysis, we first identified the haloes using \textsc{rockstar} \citep{behroozi13}, and then assembled their merger trees (also checking the consistency of the identified haloes throughout cosmic history using \textsc{consistent trees} \citep{behroozi11}. 
Then, we start from the highest redshift available, and follow the cosmic evolution of all haloes identified. This approach, which mirrors that employed in semi-analytic models, allows us to consistently follow the evolution of all haloes, and check at every step whether they fulfill a defined set of criteria for direct collapse that we next describe.

\subsection{Dynamical heating}
During the cosmic history, haloes grow via accretion from the cosmic web and merger of smaller substructures, with baryons following the dark matter and accumulating within the haloes. As the halo grows in mass with redshift $z$, gas heats up to the halo virial temperature 
\begin{equation}
    T_{\rm vir} = 1.98\times 10^4\left(\frac{\mu}{0.6}\right)\left(\frac{\Omega_{\rm m}}{\Omega^z_{\rm m}}\frac{\Delta_{\rm vir}}{18\upi^2}\right)^{1/3}\left(\frac{hM_{\rm vir}}{10^8\msun}\right)^{2/3}\frac{1+z}{10},
\end{equation}
where $\mu=1.22$ is the mean molecular weight for a fully neutral medium, $\Omega^z_{\rm m} = \Omega_{\rm m}(1+z)^3/[\Omega_{\rm m}(1+z)^3+\Omega_\Lambda]$, $\Delta_{\rm vir}= 18\upi^2+82d-39d^2$, $d=\Omega_{\rm m}^z-1$, and $M_{\rm vir}$ is the halo virial mass in solar units~\citep{barkana01}.

Given the primordial composition, baryons can cool only via hydrogen and helium atomic cooling (above $T\sim 10^4$~K) and via molecular hydrogen cooling at lower temperatures. If we define the H$_2$ cooling time 
\begin{equation}
    \tau_{\rm H_2} = \frac{1}{\gamma-1}\frac{n_{\rm gas}k_{\rm B}T}{\Lambda_{\rm H_2}n_{\rm H}^2f_{\rm H_2}},
\end{equation}
where $n_{\rm gas}=\rho/(\mu m_{\rm H})$, $\rho$ is the gas density, $k_{\rm B}$ is the Boltzmann constant, $m_{\rm H}$ is the hydrogen mass, $n_{\rm H}=0.76\rho/m_{\rm H}$ is the hydrogen nuclei density, $\gamma=5/3$ is the adiabatic index, $f_{\rm H_2}=n_{\rm H_2}/n_{\rm H}$ is the molecular hydrogen number fraction, and $\Lambda_{\rm H_2}$ is the H$_2$ cooling rate, which can be approximated as $10^{-27.6}(T_{\rm vir}/100)^{3.4}\rm\, erg\, s^{-1}\, cm^3$ for $120~{\rm K}\leq T\leq 6400~{\rm K}$ \citep{trenti09}, and compare it with the Hubble time in the matter-dominated era
\begin{equation}
    \tau_{\rm H} \approx \frac{2}{3H_0\Omega_{\rm m}^{1/2}}(1+z)^{-3/2},
\end{equation}
we obtain an estimate for the minimum H$_2$ fraction required for efficient cooling. If we assume the presence of a non-negligible Lyman-Werner background $F_{\rm LW}=4\upi J_{21}10^{-21}\rm\, erg\, s^{-1}\, cm^{-2}\, Hz^{-1}$, we can write the equilibrium H$_2$ fraction able to form in the halo as
\begin{equation}
    f_{\rm H_2} = \frac{k_{\rm H-} n_{\rm e}}{k_{\rm diss}},
\end{equation}
where $k_{\rm H-} = 1.8\times 10^{-18} T^{0.88}\rm\, cm^3\, s^{-1}$ is the coefficient for the reaction H+e$^-\rightarrow$~H$^-$+$\gamma$, assumed to be the dominant process and the bottleneck in H$_2$ formation, $n_e=x_e n_{\rm H}$ is the electron density, and 
$k_{\rm diss}=F_{\rm LW}/9\times 10^{-9} \rm\, s^{-1} = J_{21}/7.16\times 10^{11}\rm\, s^{-1}$. For $x_{\rm e}$, we assume the relic abundance $x_{\rm e} = 1.2\times 10^{-5} \sqrt{\Omega_{\rm m}}/(\Omega_{\rm b}h)$ \citep{peebles93}.

By equating this equilibrium fraction to the minimum required for cooling, and assuming $T=T_{\rm vir}$, $\rho=\eta_{\rm gas} \rho_{\rm max}$, with $\rho_{\rm max} \equiv 187\mu m_{\rm H}\Omega_{\rm b}h (T_{\rm vir}/1000~{\rm K})^{1.5}$ the maximum gas density reachable by adiabatic contraction \citep{fernandez14} and $0\leq\eta_{\rm gas}\leq 1$ an arbitrary scaling factor, we can derive the minimum halo mass as
\begin{equation}
M_{\rm vir,min}^b (\msun) \approx \frac{58.873}{\alpha_{\rm vir}^{1.5}}\xi_{21}^{0.239}n_{\rm H,0}^{-0.478}(1+z)^{0.358},
\label{eq:mmin}
\end{equation}
where $\alpha_{\rm vir} = T_{\rm vir}/ (M_{\rm vir}/\msun)^{2/3}$, $\xi_{21} = J_{21}(h\sqrt{\Omega_{\rm m}})/(\mu x_{\rm e})$, 
and $n_{\rm H,0} = 0.76\eta_{\rm gas}(\rho_{\rm max}/m_{\rm H})(T_{\rm vir}/1\,{\rm K})^{-1.5}$.

\begin{figure}
    \centering
    \includegraphics[width=\columnwidth,trim=0.5cm 0.5cm 2cm 1.5cm,clip]{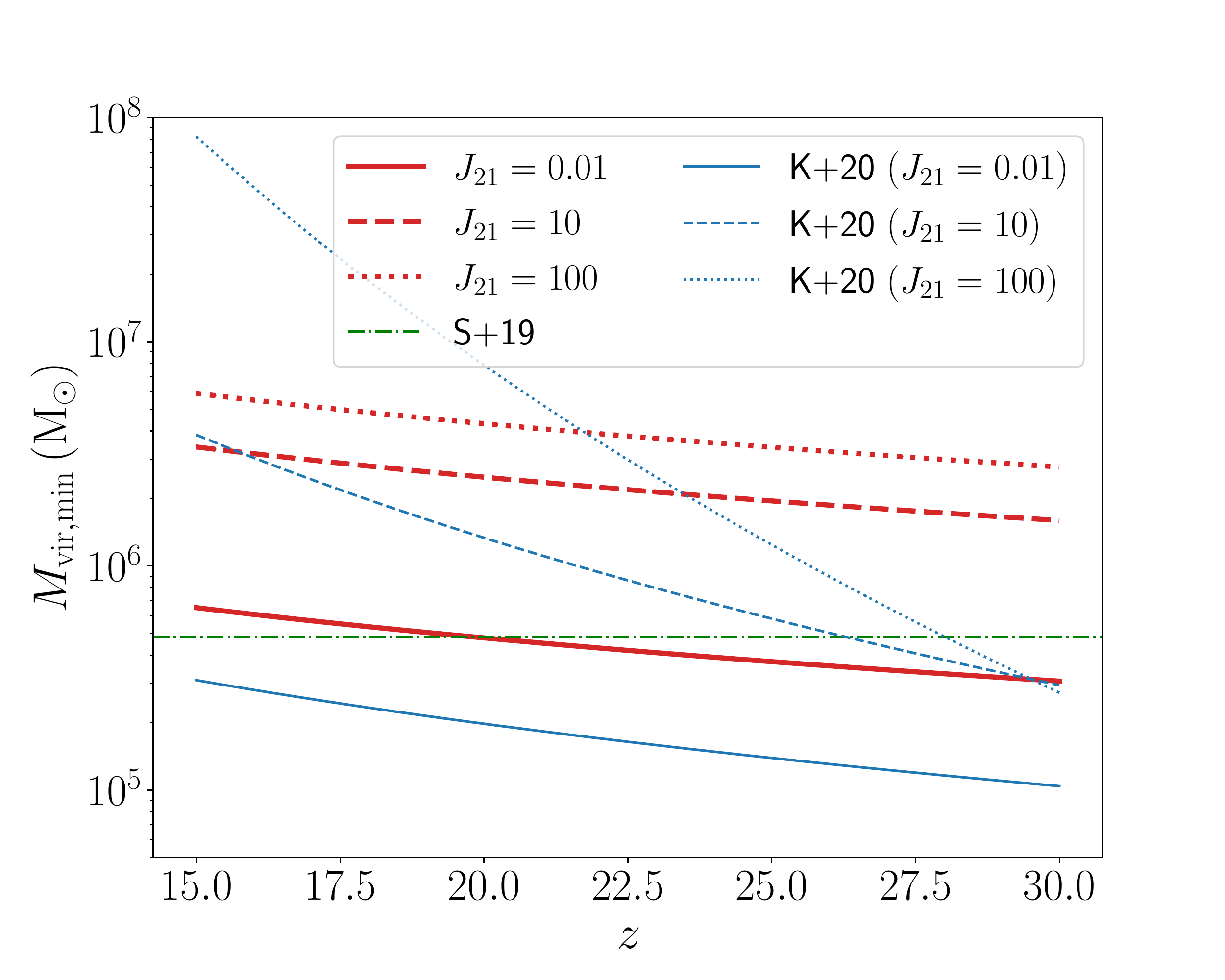}
    \caption{Minimum mass for efficient H$_2$ cooling from Eq.~\eqref{eq:mmin} for different values of $J_{21}$ (red lines), compared with the results by \citet{kulkarni20}, shown as blue lines, and the redshift-independent result (for $J_{21}=0$) by \citet{schauer19}.}
    \label{fig:mmin}
\end{figure}

Recently, \citet{schauer19} and \citet{kulkarni20} inferred the minimum mass for H$_2$ cooling from hydrodynamic simulations (performed with different codes, setups and sub-grid prescriptions), finding significant discrepancies. Hence, it is important to check how our Eq.~\eqref{eq:mmin}, based on simple scaling relations and an approximate form of the H$_2$ cooling function, compares to these numerically derived results. In Fig.~\ref{fig:mmin}, we show a comparison of the minimum masses over the redshift range $z=15-30$ for different values of $J_{21}$, i.e. 0.01 (solid lines), 1 (dashed lines), and 10 (dotted lines). The results by \citet{kulkarni20} are shown in blue (K+20), our Eq.~\eqref{eq:mmin} with $\eta_{\rm gas}=0.2$ in red,
and the redshift-independent result by \citet{schauer19} (with $J_{21}=0$) as a green dot-dashed line (S+19). From the figure, it is clear that our results exhibit a decreasing trend with redshift similar to \citet{kulkarni20}, although with a higher normalisation resulting from $\eta_{\rm gas}=0.2$. Indeed, if $\eta_{\rm gas}=1$ is considered, the mass decreases by a factor of $\sim 2$, getting closer to \citet{kulkarni20}. Nevertheless, the difference always remains to within a factor of $\sim$ three over most of the relevant redshift range. On the other hand, the results by \citep{schauer19} always remain in between, but are higher for the case of no impinging LW flux. 

To conclude, our Eq.~\eqref{eq:mmin} represents a reasonable estimate, in rough agreement with state-of-the-art numerical results over the relevant ranges of redshift and $J_{21}$. Nonetheless, the minimum masses we find are typically a factor of $\sim 2-3$ higher than those found in \citet{kulkarni20}, except below $z\lsim 16$ and above $J_{21}\gsim 10$, where our threshold is lower than their fit. In principle, lowering the mass thresholds could allow cooling in additional low-mass mini-haloes and reduce the number of DCBH host candidates, while increasing the threshold could have the opposite effect below $z\sim 16$.  To address this caveat, we have re-run our models  using the fitting formulae from \citet{kulkarni20}. We found that the differences in our results were almost negligible at very high redshift, although below $z=15$, it resulted in a few percent higher number of candidates relative to our fiducial case. However, we caution that our ability to fully test this caveat is limited by the resolution of our simulations (with a particle mass of $M_{\rm dm}\approx 9\times 10^4\,\msun$, we are unable to resolve mini-haloes with masses much below $10^6\,\msun$).

Above this mass threshold, a plausible process that is able to counteract radiative cooling is the compressional heating provided by the mass accretion onto the halo. This has been dubbed "dynamical heating" in the literature, 
and the corresponding heating rate $\Gamma_{\rm dyn}$ can be approximated as 
\begin{equation}
    n\Gamma_{\rm dyn} = \frac{n_{\rm H}}{0.76\mu}\frac{k_{\rm B}}{\gamma-1}\left(\frac{2\alpha_{\rm vir}}{3}M_{\rm vir}^{-1/3}\frac{\de M_{\rm vir}}{\de t} + \frac{T_{\rm vir}}{1+z}\frac{{\rm d}z}{{\rm d}t}\right),
    \label{eq:dynheat}
\end{equation}
where the first term on the right-hand side represents compressional heating accompanying the mass growth of the halo \citep{yoshida03}, and the second term accommodates the decrease of the characteristic densities and the virial temperature (at fixed halo mass) towards lower redshift, due to the expansion of the Universe. Numerically, $\Gamma_{\rm dyn}$ is computed by finite difference between halo masses from consecutive snapshots.

In order to prevent H$_2$ cooling, we need $n\Gamma_{\rm dyn}\geq f_{\rm H_2}n_{\rm H}^2\Lambda_{\rm H_2}$. Since, for short enough time-scales (order of Myr at the high redshifts considered in this work), the second term on the right-hand side of Eq.~\eqref{eq:dynheat} is small, it can be neglected, translating the condition into a critical mass accretion rate
\begin{equation}
    \left(\frac{\de M_{\rm vir}}{\de t}\right)_{\rm crit}= 0.76\mu f_{\rm H_2} n_{\rm H}\frac{3(\gamma-1)}{2k_{\rm B}}\alpha_{\rm vir}^{-1}\Lambda_{\rm H_2}M_{\rm vir}^{1/3}.
\end{equation}

We follow the evolution of all mini-haloes in the simulation, and track whether at any stage they exceed the minimum mass for H$_2$ cooling.   For any mini-halo that exceeds this threshold, we further check whether they fulfill the dynamical-heating criterion.  For any halo above the H$_2$ cooling minimum mass which, during any step of the evolution, fails the latter criterion, we assume that it remains a plausible host of a direct-collapse BH only for one H$_2$ cooling time $\tau_{\rm H_2}$~\citep{fernandez14}, after which it is assumed to form stars and is excluded from the candidate list, unless the condition for dynamical heating was fulfilled again within the time $\tau_{\rm H_2}$. 
For example, a halo could experience a strong accretion or a merger at a given time, events that result in a dynamical-heating rate much higher that the critical value, then the accretion drops below the threshold for a short time, and then increases again above threshold. If the time during which the halo is below the critical threshold is shorter than the halo H$_2$ cooling time $\tau_{\rm H_2}$, we retain the halo as a candidate.

\subsection{Synchronised pairs}

While dynamical heating may prevent H$_2$ cooling and star formation in the mini-halo stage, it is unclear whether it can help avoid catastrophic cooling (via H atoms and then H$_2$) and the ensuing star formation once a halo crosses the atomic-cooling threshold (ACT), even if the halo is pristine (see more discussion of this below). A plausible scenario proposed to achieve this, and keep even ACHs free of H$_2$ cooling, involves synchronised pairs of ACHs. 
In this scenario, once the first halo crosses the atomic cooling threshold (Halo 1), it cools and forms stars in $\Delta t_{\rm SF} = t_{\rm ff} \equiv \sqrt{(3\upi)/(32{\rm G}\rho)}$, subsequently providing a strong enough LW flux able to prevent the formation of H$_2$ in the companion halo (Halo 2). This allows gas in Halo 2 to collapse nearly isothermally at $T\sim 10^4$~K, leading to the conditions (i.e. large accretion rate) required for the formation of a DCBH. 

In order for this process to occur, in our model we require the time difference $\Delta t$ between the onset of SF in Halo 1, occurring at $t=\tau_{\rm ff,1}$ after the ACT crossing, and the ACT crossing in Halo 2 to be shorter than $\Delta t_{\rm SN}=5$~Myr, the characteristic time after which the first massive stars explode, ejecting metals that enrich the medium~\citep[see, e.g.][]{regan17}. This translates to a maximum time separation between the two ACT crossing events $\tau_{\rm ff,1}<\Delta t < \tau_{\rm ff,1} + \Delta t_{\rm SN}$. In addition, to ensure that the LW flux is large enough to prevent H$_2$ formation, we also require the halo separation to be $200 < \Delta r < 500$~physical pc. 
Since, in our model, we also explore the case of star formation occurring in mini-haloes that had not been sufficiently dynamically heated, in these cases we exclude from the list of synchronised pair candidates all haloes that had formed stars during their mini-halo phase. Nevertheless, below we also compare such results with a simpler approach based on the number of progenitors, as discussed in Section~\ref{sec:pollution}.

\subsection{The LW radiation effect}

An important role in determining the abundance of potential DCBHs is played by the value of $J_{21}$ chosen for the analysis. In this work, we consider three different cases. In the first two, $J_{21}$ is kept constant at 0.01 and 1.0, respectively, which brackets the fiducial values in \citet{trenti09}, and values expected for the mean LW background at the redshifts considered here. In the third, we self-consistently compute the local $J_{21}$ due to nearby star-forming haloes, which can be much larger than the background, depending on the halo's location. We follow \citet{dijkstra14} and compute the local flux impinging on each halo as the sum of the fluxes over all star-forming haloes in the tracked region, with the contribution of each halo given by
\begin{equation}
\label{eq:j21}
    J_{21}=\frac{h\langle\nu\rangle}{\Delta\nu}\frac{f_{\rm esc,LW}Q_{\rm LW}}{16\upi r^2}M_\star,
\end{equation}
where $M_\star=f_\star(\Omega_{\rm b}/\Omega_{\rm m})M_{\rm vir}$.
Here we use $\langle\nu\rangle=2.99\times 10^{15}$~Hz and $\Delta\nu=5.79\times 10^{14}$~Hz as the average frequency and the bandwidth of the LW band,
extending from 11.2 to 13.6~eV,
and $f_{\rm esc,LW}$ is the escape fraction of LW radiation from the halo, which we set to 100 per cent. The latter assumption is reasonable in most cases, in particular far away from the irradiating halo, and when the relative gas velocities are large, hence doppler-shifting the LW photons out of resonance.  On the other hand, the measured LW $f_{\rm esc}$ can drop significantly closer to the halo, if the stellar ionisation front  expands slowly and/or if the stellar spectrum softens, allowing for H$_2$ to form in the outskirts of the halo and absorb LW radiation \citep[see, e.g.,][]{schauer17}. 
Finally, $r$ is the distance from the LW source, and 
\begin{equation}
    Q_{\rm LW} = Q_0[1+t_{\rm LW}/(4{\rm\, Myr})]^{-3/2}\exp[-t_{\rm LW}/(300\,\rm Myr)]
    \label{eq:qlw}
\end{equation}
is the LW photon production rate per solar mass formed in stars per single burst of SF, at a time $t_{\rm LW}$ after a SF burst, with $Q_0$ the normalisation, which we set to $10^{47}\rm\, s^{-1}\, \msun^{-1}$ for both PopIII and PopII stars.  We adopt the photon yields from \citet{schaerer03} and the same initial mass function for both populations, in order to keep the same functional form for the time dependence of the LW flux. The decaying trend of $Q_{\rm LW}$ reflects the fact that massive stars, that are the main contributors to the LW radiation field, live only for a few Myr. 
Before discussing the redshift evolution, we need to point out a possible limitation of this treatment of the LW radiation. Eq.~\eqref{eq:j21} is self-consistent only in the case of SF occurring in a single burst and then instantaneously stopping. If instead SF occurs continuously, as one would expect, $J_{21}$ should be determined by the convolution of $Q_{\rm LW}$ with the time-dependent star formation rate (SFR) in the halo. 
Unfortunately, such a treatment cannot be easily implemented in a semi-analytic framework, hence we opt here for an approximation in which $Q_{\rm LW}$ in each halo is evolved at every time-step $\Delta t$ as
\begin{equation}
    Q_{\rm LW}' \approx Q_{\rm LW}[1+\Big\langle\frac{\dot{Q}_{\rm LW}}{Q_{\rm LW}}\Big\rangle'\Delta t]\frac{M_\star}{M'_\star},
    \label{eq:qprime}
\end{equation}
where the primed quantities correspond to the updated value at $t_{\rm LW}+\Delta t$,
\begin{equation}
    \Big\langle\frac{\dot{Q}_{\rm LW}}{Q_{\rm LW}}\Big\rangle' \approx -\frac{1}{300}-\frac{3}{8}\left(1+\frac{t'_{\rm LW}-\langle\tau'_{\rm LW}\rangle}{4\,\rm Myr}\right)^{-1}
\end{equation}
is the average decay rate of the LW flux of the entire population, and
\begin{equation}
    \langle\tau_{\rm LW}\rangle' = \frac{\langle\tau_{\rm LW}\rangle M_\star Q_{\rm LW} + t_{\rm LW}'(M'_\star-M_\star)Q_0}{M_\star Q_{\rm LW}+(M'_\star-M_\star)Q_0}
    \label{eq:tauprime}
\end{equation}
is the flux-weighted average time of the SF bursts at $t'$. In order to accurately follow the variations of $Q_{\rm LW}$, we also need to impose a time-step limiter to the integration, which we define as $\Delta t = \min\{0.1|\langle Q_{\rm LW}/\dot{Q}_{\rm LW}\rangle|,\Delta t_{\rm sim}\}$. If the integration time-step is shorter than the time interval between the simulation snapshots, we sub-cycle Eqs.~\eqref{eq:qprime} to \eqref{eq:tauprime} assuming $M'_\star = M_\star + \Delta M_{\star,\rm sim}(\Delta t/\Delta t_{\rm sim})$, with $\Delta M_{\star,\rm sim}=M_\star[t+\Delta t_{\rm sim}-M_\star(t)]$.
It can be easily seen that our approximation reduces to the exact solution for a single burst, in which $M_\star'=M_\star$ yields $\langle \tau_{\rm LW}\rangle' = \langle\tau_{\rm LW}\rangle$, and Eq.~\eqref{eq:qprime} reduces to an explicit time integration of the $Q_{\rm LW}$ decay with time. On the other hand, in the case of a  constant SFR with $M_\star'\gg M_\star$, Eq.~\eqref{eq:tauprime} reduces to $\langle\tau_{\rm LW}\rangle'\lesssim t_{\rm LW}'$, that gives an almost exact solution in which $Q_{\rm LW}$ slowly decays with time according to the mass ratio between young stars and old stars. Although not explicitly shown, we verified the accuracy of our approximation in different cases, finding an extremely good agreement always within $\sim 10\%$.

In the variable $J_{21}$ case, we always enforce a minimum $J_{21,\rm min}=0.01$ to be present in the box, aimed at mimicking a global background from sources outside the box \citep{trenti09}. Such a floor also avoids the divergence in the H$_2$ fraction calculations, which should be changed accounting for collisional H$_2$ dissociation in the absence of radiation.

\subsection{Redshift evolution}

The criteria described above are applied to every output of the simulation, starting from the highest redshift down to $z=10$, similar to the procedures in semi-analytic models. Although the results for synchronised pairs and dynamically heated haloes are presented in different Figures, they are tracked simultaneously in the code, i.e., they can co-exist in the same volume for a given set of parameters.

In particular, our procedure is the following:
\begin{enumerate}
    \item We read a new \textsc{rockstar} output containing the list of haloes to add and evolve.
    \item If a halo progenitor is not present (the current halo has been identified for the first time), we add it to the list of haloes to evolve. If its virial temperature is below $10^4$~K, the halo is considered to be a candidate for becoming a DCBH host, according to the criteria outlined above. Otherwise (this only occurs when the halo forms at the boundary of the high-resolution region, hence it is contaminated by massive particles and the mini-halo phase is not resolved) it is conservatively considered as star-forming. These star-forming haloes are excluded from the list of DCBH host candidates, but are a potential source of metal contamination (see below).
    \item If a progenitor is present, and is a mini-halo, we determine the halo accretion rate during the last step and check the dynamical heating criterion. If the criterion is not matched, the halo is removed from the candidate list, but evolved further as a possible source of contamination. If the target halo has just crossed the ACT, and the dynamical heating criterion has always been matched, we flag it as dynamically heated and store it in a dedicated catalogue of candidates.
 
    \item If, on the other hand, the progenitor was already above the ACT, the halo is simply updated, accounting for star formation and metal enrichment, the latter computed according to \citet{dijkstra14}.
    \item among all updated haloes, we identify those that have just crossed the ACT, and look for nearby haloes matching the synchronisation criteria described above. Because of tidal stripping during halo mergers, it can occur that some haloes keep oscillating around the ACT for some time after their first crossing. In order to avoid our prescription to spuriously identify pairs multiple times, we remove the pairs from the candidate list as soon as they have crossed the ACT for the first time. Also for synchronised pairs, we store the pair properties in a catalogue.
    \item We then repeat the above for all haloes, checking each halo for metal pollution inhibiting the formation of DCBHs (see next subsection);
    \item Finally, before moving to the next snapshot, we check for mergers among the haloes, identifying those haloes sharing the same descendant; then, we update the list of active haloes by creating a unique virtual descendant for each merger, represented by its most massive progenitor; we then consistently update the number of progenitors of this virtual descendant, and
    also take into account whether any of the progenitors had already crossed the ACT in the past or was contaminated with metals.
\end{enumerate}

\subsection{Star formation and metal pollution}
\label{sec:pollution}

An important process that can affect our results is the possible contamination by metals, via genetic pollution (when progenitors already had metals) and via environmental pollution (when the metal bubbles created after SN explosions in nearby galaxies reach the target halo). In our model, we consider both of these processes. 

For genetic pollution, in this work we consider two different models. A first approach consists of tracking only the number of progenitors $N_{\rm prog}$ of each candidate halo, and, assuming that each progenitor has a fixed 10\% probability of becoming star-forming (SF) during its history \citep{dijkstra14}.  The advantage of this simple treatment is that SF in mini-haloes does not need to be followed explicitly. Instead, we filter the candidate list by applying two different criteria: in order to avoid metal-pollution, we require either $N_{\rm prog}<2$ or $N_{\rm prog}<5$, which correspond to at most 10\% or 40\% probability of being star forming during the mini-halo phase, respectively.  We obtained results from this simple approach as an academic exercise, and in order to be able to compare results with previous work by \citet{dijkstra14}. In the second approach, instead, which we employ as our fiducial case, we self-consistently account for SF in mini-haloes that can cool via H$_2$, and exclude these from the list of candidate dynamically heated mini-haloes.

For environmental pollution, we follow all haloes able to form stars in our cosmological volume, i.e. mini-haloes not dynamically heated and all ACHs, and compute the size of the metal bubble starting from the time of the first SN explosion. Hence, the metal bubbles start to expand in our model $\Delta t = t_{\rm ff}+\Delta t_{\rm SN}$ after the halo is first declared star-forming.

To model SF, we assume that the stellar mass formed in ACHs is $f_\star=0.05$ of the baryonic mass, translating to $M_\star = f_\star \Omega_{\rm b}/\Omega_{\rm m}M_{\rm vir}$. For mini-haloes, instead, we consider a lower star-forming efficiency $f_\star=0.005$.
In both cases, we assume that SF starts with a burst as soon as one free-fall time (defined at the maximum density achievable via adiabatic contraction $\rho_{\rm gas}=\eta_{\rm gas}\rho_{\rm max}$) has elapsed since (i) dynamical heating conditions have failed or (ii) the ACT has been exceeded (if no previous SF in the mini-halo had already occurred), and then proceeds smoothly following the increase in stellar mass with time in the halo.
Following \citet{weaver77} and \citet{madau01b}, we define the bubble radius as
\begin{equation}
    R_{\rm bubble}(t) = \left(\frac{125}{154\upi}\right)^{1/5}\left(\frac{L_{\rm SN}}{{n \rm m_H}}\right)^{1/5}t^{3/5} \rm\, cm,
\end{equation}
where $L=\rm const$ is the average SN luminosity associated to the stellar population in the halo, $t$ the time since the burst of SF, and $n \equiv \Delta_{\rm gas}\Omega_{\rm b}\rho_{\rm crit}(1+z)^3/m_{\rm H}$ is the gas density (in cm$^{-3}$) the bubble expands through.
Now, we assume for simplicity that $L_{\rm SN}\approx E_{\rm SN}\eta_{\rm SN}M_\star(t)/t$, where $E_{\rm SN}=10^{51}\rm\, erg$ is the energy released per SN, $M_\star(t) = f_\star (\Omega_{\rm b}/\Omega_{\rm m}) M_{\rm vir}(t)$ is the stellar mass at time $t$, and $\eta_{\rm SN}\approx 0.011\, \msun^{-1}$ is the number of SNe per stellar mass formed, and replace them in the bubble radius equation, we obtain, for $t>\Delta t_{\rm SN}$,
\begin{equation}
    R_{\rm bubble}(t) \approx 23\,{\rm pc} \left[f_\star\frac{\Omega_{\rm b}}{\Omega_{\rm m}}\frac{M_{\rm vir}(t)}{\msun}\frac{\rm Myr}{t}\left(\frac{t-\Delta t_{\rm SN}}{\rm Myr}\right)^3\right]^{1/5}n^{-1/5}.
    \label{eq:rbub}
\end{equation}
We notice that this equation reduces, for $t\gg \Delta t_{\rm SN}$, to the equation reported in \citet{dijkstra14}, apart from the slightly lower normalisation. 
In \citet{dijkstra14}, the baryon overdensity through which the bubble expands is assumed to be $\Delta_{\rm gas}\approx 60$, typical of the overdensities found within haloes ($60\lesssim \Delta_{\rm gas} \lesssim 1000$), although the bubble spends most of the time in the inter-galactic medium at $\Delta_{\rm gas}\sim 1-10$. This results in the bubble size being underestimated \citep{dijkstra14,habouzit17}. In our model, instead, we also account how the overdensity changes outside the halo, by fitting the asymptotic case of an extremely rare halo with $M_{\rm vir}=10^8\,\msun$ from \citet{barkana04}, and simply express $\Delta_{\rm gas}$ as
\begin{equation}
    \Delta_{\rm gas} =
    \begin{cases}
        60 & R\leq 0.787R_{\rm vir}\\
        10^{-0.0686x^3+0.5664x^2-1.6088x+1.6051} & \rm otherwise,
    \end{cases}
\end{equation}
where $x\equiv\log_{10}(R/R_{\rm vir})$, $R$ is the bubble radius, and the factor 0.787 has been chosen to join the inner constant value and the extrapolation of the fitting function without discontinuity.
To consistently find the bubble radius outside $0.787R_{\rm vir}$, when $\Delta_{\rm gas}$ also depends on $R_{\rm bubble}$, we iterate until convergence.
At every step, we flag all candidate haloes that are enriched by metal bubbles, by checking whether the bubble overlaps with the halo, i.e. the halo separation $r<R_{\rm bubble}-R_{\rm vir, candidate}$. A possible caveat in this evolution comes from the stellar mass scaling in the bubble radius equation, which does not consistently keep track of the superposition of metal bubbles from multiple SF events in the same halo, and might result in either an underestimate or an overestimate of the actual bubble size, depending on how the bubbles interact.

\section{Results}
\label{sec:results}

In this section, we assess the importance of the two mechanisms discussed above for the formation of DCBH host candidates. We first focus on the synchronised-pair scenario, and then discuss the role of dynamical heating, but we again stress that we follow both at the same time in our models.
A list of the cases we explored, with the differences in their assumptions, is summarised in Table~\ref{tab:cases}. In particular, while in model `Visbal'  we assume the same SF and synchronisation parameters as in \citet{visbal14b}, all other models employ different SF-synchronisation criteria, i.e. we assume that SF occurs one free-fall time after the halo has been declared eligible for SF, with the free-fall time estimated at a density $\rho_{\rm gas}=\eta_{\rm gas}\rho_{\rm max}$. For our fiducial model, we assume $\eta_{\rm gas}=0.2$, consistent with the results in \citet{visbal14a}, and explore the case of $\eta_{\rm gas}=1$ only in our HighDensity model.

\begin{table}
    \centering
    \caption{List of the models explored, with the differences in their assumptions. The first column is the name of the model, the second the time delay between when a halo is declared star-forming and the formation of the first stellar population, and the third is the time delay before the first SNe explode. The fourth column reports the density in units of the maximum density achievable via adiabatic contraction and the fifth the assumed value of the LW flux.
    The first two models (top of the table) exclude SF in mini-haloes, while the other four models (bottom of the table) include SF in the subset of mini-haloes which can cool, according to the criteria discussed in the text.}
    \begin{tabular}{lcccc}
    \hline
        Model name & $\Delta t_{\rm SF}$ & $\Delta t_{\rm SN}$ & $\eta_{\rm gas}$ & $J_{21}$\\
        \hline 
        \multicolumn{5}{c}{Basic models: no SF in mini-haloes}\\
        \hline
        Visbal &  10~Myr & 10~Myr & - & - \\
        NoMini & $t_{\rm ff}(n_{\rm gas})$ & 5~Myr & 0.2 & - \\
        \hline 
        \multicolumn{5}{c}{Full models with SF in mini-haloes}\\
        \hline
        Fiducial & $t_{\rm ff}(n_{\rm gas})$ & 5~Myr & 0.2 & 0.01+Local \\
        LowFlux & $t_{\rm ff}(n_{\rm gas})$ & 5~Myr & 0.2 & 0.01 \\
        HighFlux &$t_{\rm ff}(n_{\rm gas})$ & 5~Myr & 0.2 & 1 \\
        HighDensity & $t_{\rm ff}(n_{\rm gas})$ & 5~Myr & 1 & 0.01+Local \\
        
        \hline
    \end{tabular}
    \label{tab:cases}
\end{table}

\subsection{The synchronised pair scenario}

\begin{figure}
    \centering
    \includegraphics[width=\columnwidth,trim=1cm 1cm 1cm 2cm,clip]{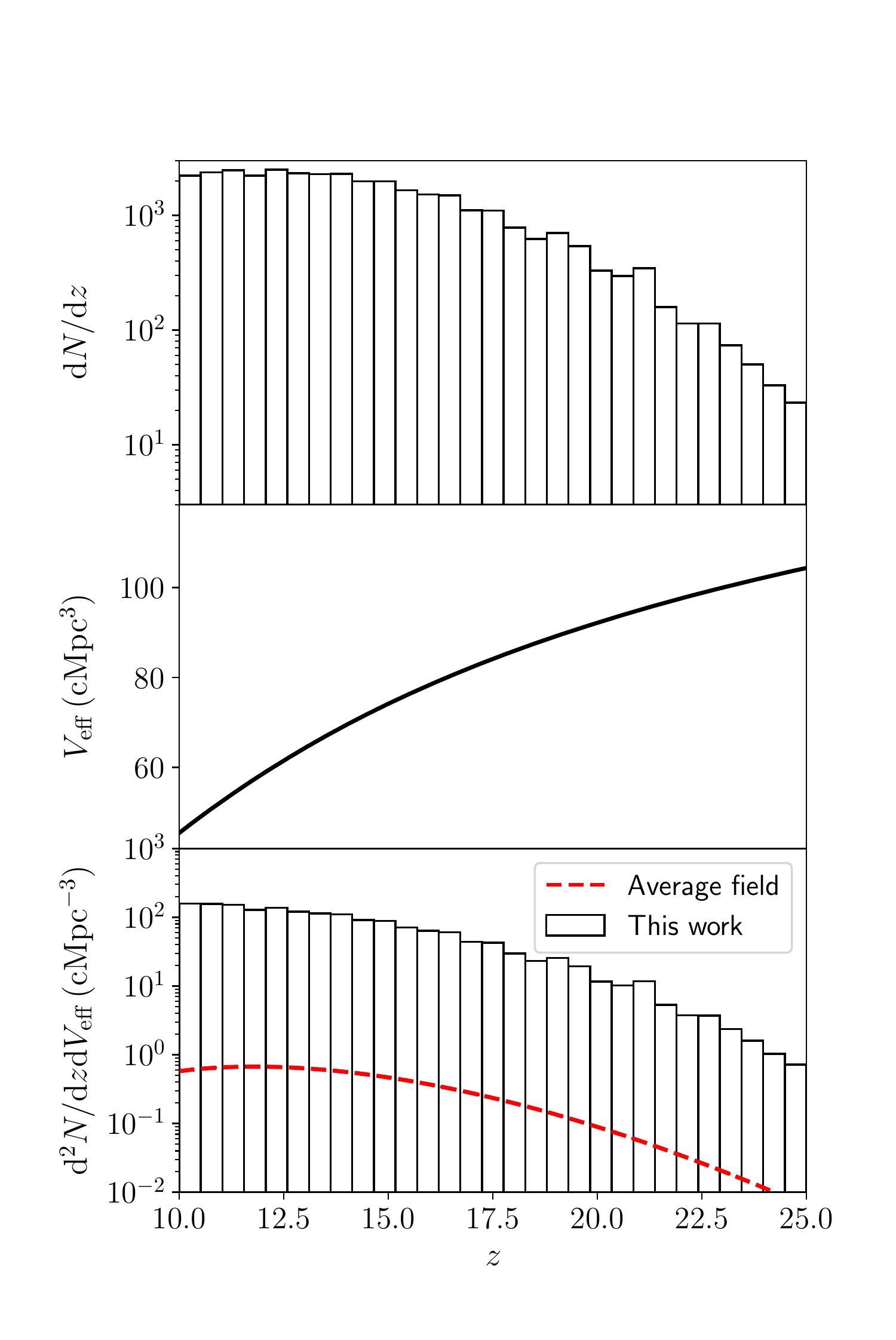}
    \caption{Redshift distribution of the haloes crossing the ACT 
   in our effective Lagrangian volume (top panel).  The effective volume is $\approx 150$ comoving Mpc$^3$ at z=100, but subsequently shrinks as the region collapses and as a smaller fraction is tracked at high resolution (middle panel). 
   The bottom panel shows the formation rate of ACHs per unit volume in our simulation (black histogram, compared to the average ACH formation rate density in the background Universe (obtained from \citealt{watson13}; red dashed curve).
   Overall, the number density of ACHs in our volume is about a factor of 100 higher than in an average region of the Universe.}
    \label{fig:zdist}
\end{figure}

In order to disentangle the effect of the large-scale overdensity from the additional physical processes included in our model, we first consider a simplified model, which we dub `Visbal', in which we follow the prescriptions employed in \citet{visbal14b},\footnote{We note that the simulations in \citet{visbal14b} employed $\sigma_8 = 0.83$, which is slightly higher than ours ($\sigma_8=0.8159$), and this could have a moderate effect on the candidate numbers.} and later include the additional physical processes and criteria described in \S~\ref{sec:method}. 
To give an idea of the number of halo candidates in the high-resolution region of our reference simulation, we show in Fig.~\ref{fig:zdist} the redshift distribution of haloes crossing the ACT down to $z=10$ (top panel). The first few hundred haloes appear already above $z=20$, and their formation rate becomes roughly constant around $z=15$, where the DCBH seed formation scenario is typically assumed to become important \citep[see, e.g.][]{dijkstra14,valiante16}, resulting in a total of 17,500 (6,043) ACHs formed by $z=10\,(15)$.
To appreciate the relevance of these numbers, they have to be normalised by volume. In the middle panel we show the effective volume of the Universe sampled as a function of redshift, identified as a sphere with radius the mean maximum separation along $x,y$, and $z$. Compared to full box simulations, our run is a zoom-in simulation of an overdense region, in which a Lagrangian volume centred around the target halo is selected, and then evolved with time following the collapse of such region. This results in a high-resolution volume that shrinks significantly with time, from about $\sim 150$ comoving Mpc$^3$ at $z=100$ down to 8~cMpc$^3$ at $z=6$, the latter corresponding to the volume of a sphere with radius 2.5 times the virial radius of the target halo ($R\sim 69$~kpc at $z=6$). 

In order to consistently compare the abundance of ACHs in our region with an average density region of the Universe, we compute the number density of ACHs in the Universe from the Halo mass function by \citet{watson13} for haloes at the ACT at different redshifts.  \footnote{To compute the Halo mass function, we employed the public tool \textsc{colossus} \citep{diemer18}.} The comparison with our region is shown in the bottom panel, with our results as a black histogram and the average density region's as a red dashed line.  Such a large density of ACHs, especially at very high redshift, is unique to the overdense regions where $z=6$ massive haloes form, characterised by a rare $\sim 5\sigma$ peak, where our simulated region produces about a 100 times larger number of ACHs with respect to an average density one.

\subsubsection{The impact of the large-scale cosmological environment}
\label{sec:visbal}
The underlying N-body simulation used in our study represents a strongly biased region,  which evolves into an ultra-rare $\sim 10^{12}~{\rm M_\odot}$ halo by $z=6$. As a result, this region is highly overdense - by a factor of $\approx 20$ 
compared to the global average at $z=6$.  In order to assess the impact of this bias, we first compare our results with those of \citet{visbal14b}, who considered the formation of synchronised (sub)halo pairs in a typical region of the Universe in a (15 Mpc)$^3$ box.
For this analysis, we make the same assumptions as \citet{visbal14b}, i.e. \textit{(i)} star formation is suppressed in all mini-haloes, \textit{(ii)} the first halo to cross the ACT starts to form stars 10~Myr after the crossing, \textit{(iii)} the second halo must cross the ACT within 10~Myr from the SF event, and \textit{(iv)} the halo pair separation should fall within the ranges (a) [0.2-0.5]~kpc, (b) [0.2-0.75]~kpc, or (c) [0.2-1.0]~kpc.  The lower limit on the separation (0.2 kpc) corresponds to a distance within which ram-pressure and/or tidal stripping could prohibit gas collapse and DCBH formation in the target halo, while the upper limit of 0.5-1 kpc represents the uncertainty in the distance over which the critical LW flux may extend.

\begin{figure}
    \centering
    \includegraphics[width=\columnwidth,trim=1cm 2cm 1cm 1.8cm,clip]{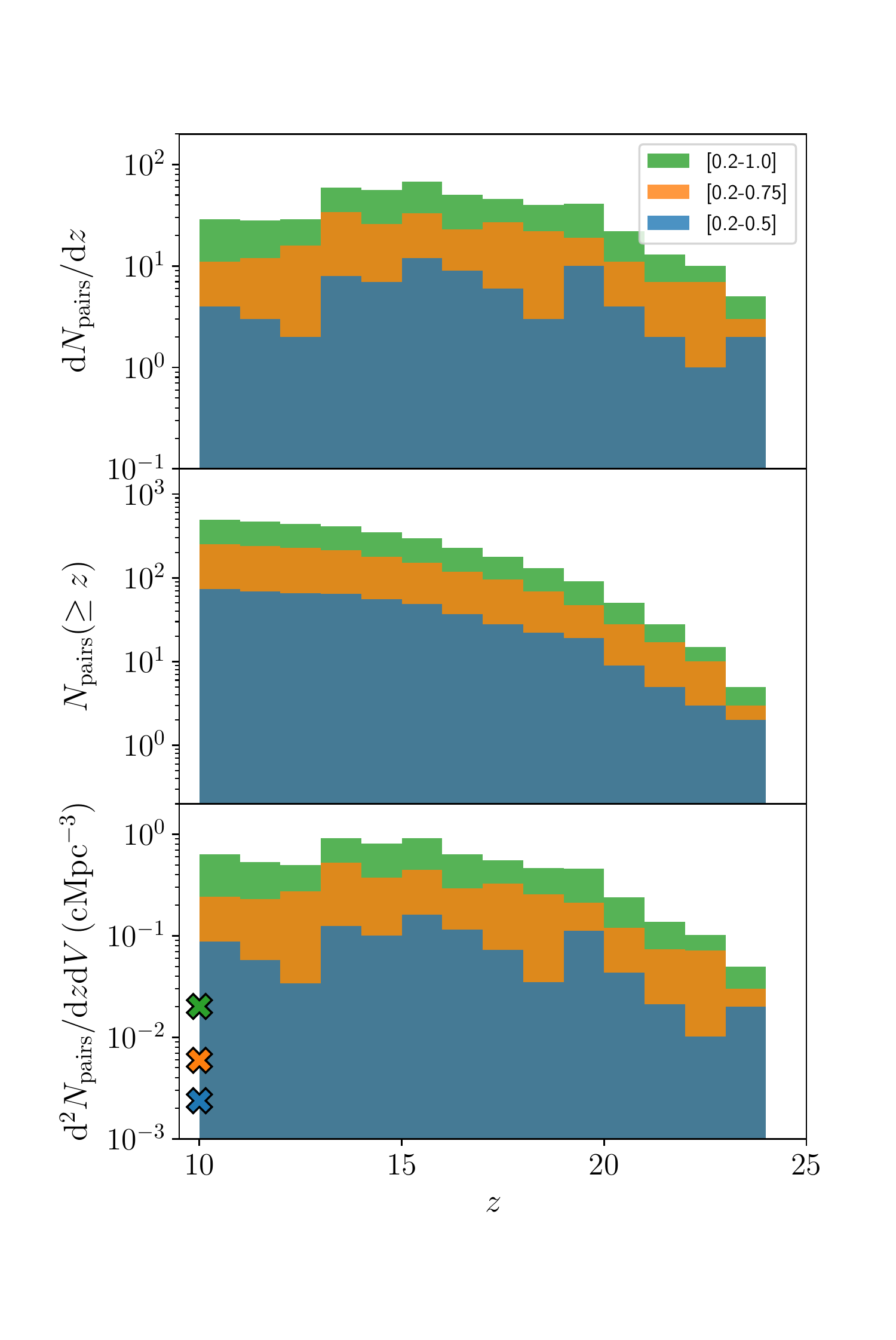}
    \caption{Redshift distribution of the 
    73, 251, and 496 
    synchronised pairs found in our biased region, following the semi-analytic prescriptions in \citet{visbal14b} for the three spatial separations between the pair members considered: [0.2-0.5] (blue histogram), [0.2-0.75] (orange histogram), and [0.2,1.0]~kpc (green histogram), respectively. The top panel shows the formation rate (per unit redshift) in the simulated volume, the middle one the cumulative number of pairs, and the bottom one the formation rate density per comoving unit volume. The three crosses with the same colour scheme show the corresponding results by \citet{visbal14b} in an average density volume of (15~cMpc)$^3$.}
    \label{fig:visbal}
\end{figure}

The results of this comparison are reported in Fig.~\ref{fig:visbal}. The top panel of this figure shows the redshift distribution, the middle one the cumulative number of synchronised pairs, and the bottom one the number of pairs per unit volume per redshift bin we found. The corresponding results by \citet{visbal14b}, obtained in a volume of (15~cMpc)$^3$ are also shown as coloured crosses in the bottom panel.
Blue, orange, and green colours correspond to the three separation intervals considered, respectively, as labelled in the legend of the figure.

The number of synchronised pairs in our simulation can reach up to a few hundred down to $z=10$ in the most optimistic case (green histogram), with a rate that varies between a few (blue histogram) and a few tens (green histogram), depending on the allowed range of separations (the smaller the maximum separation, the lower the number of candidates). Compared to the results in \citet{visbal14b}, we observe the same trend as a function of the maximum separation allowed, while the overall number of candidates in the range $z=[10-11]$ is 4, 11, and 29 for separations (a), (b), and (c), respectively.
For comparison, \citet{visbal14b} identified 2, 5, and 17 pairs in their box in the interval $\Delta z=0.25$ around $z=10$. Re-scaled to our $\Delta z=1$ and effective volume of $\approx 45$~cMpc$^3$ at $z=10$, these correspond to 0.11, 0.27, and 0.9 pairs for separations (a), (b), and (c), respectively.
This demonstrates that the overdense environment of our simulation favours the formation of such systems by a factor of $\sim 20-40$. This comparison is based on the effective volume at $z=10$, and some of the difference is due to the factor of $\sim 3$ reduction in our Lagrangian volume due to the collapse of the whole region. If we employ our initial Lagrangian volume of $150$ Mpc$^3$ at $z=100$ instead, we find that our biased region has a factor of 6--14 more synchronised pairs compared to \citet{visbal14b}. 

\subsubsection{A simplified approach to genetic and environmental pollution}
\begin{figure}
    \centering
    \includegraphics[width=\columnwidth,trim=0.5cm 3.6cm 1cm 1.8cm,clip]{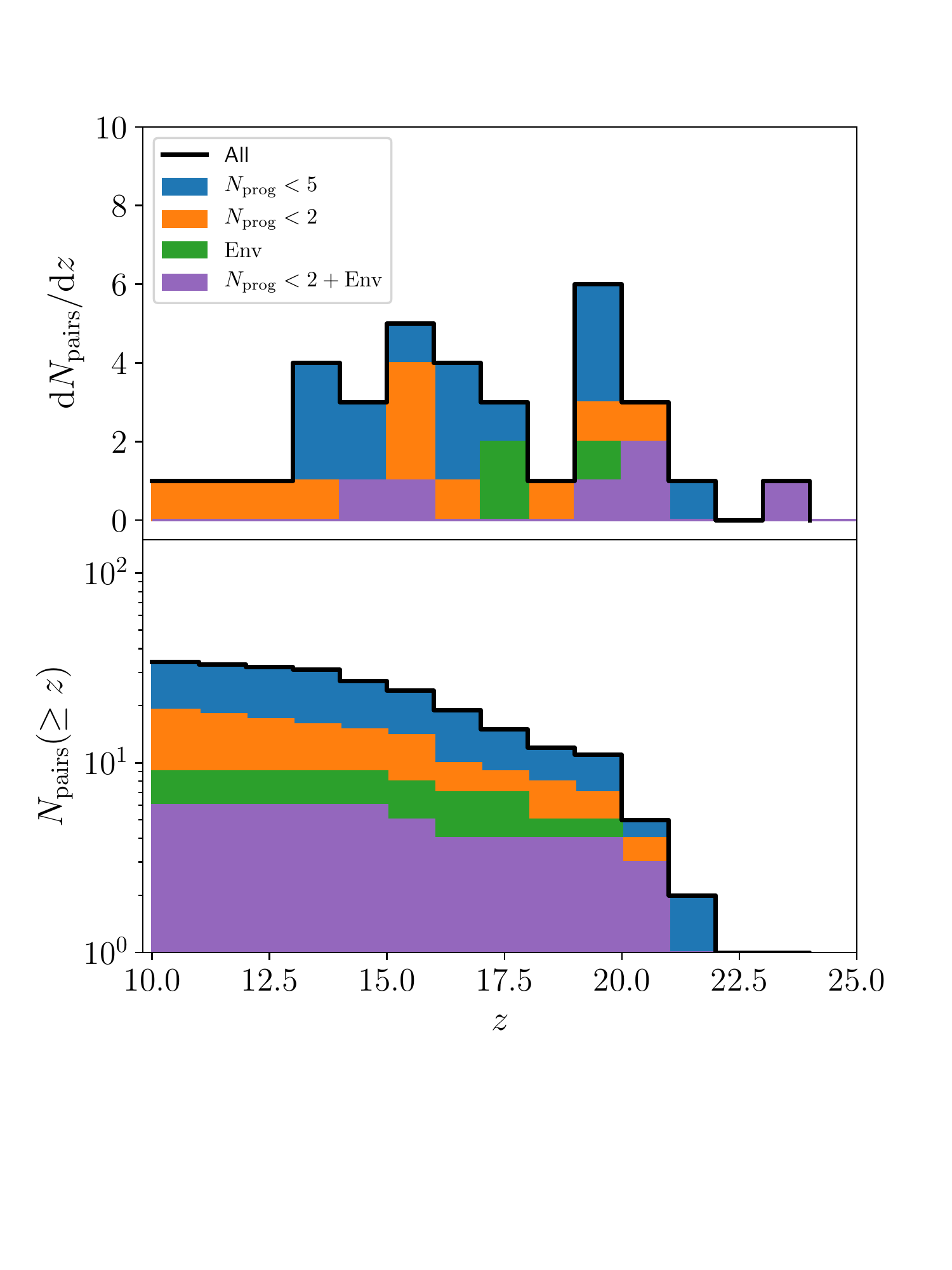}
    \caption{Redshift distribution of the 34
    synchronised pairs in our NoMini model (black histogram), 
    with the additional criteria imposed to exclude genetically (blue and orange histograms for $N_{\rm prog}<5$ and $N_{\rm prog}<2$ respectively) or environmentally (green histogram) metal-polluted haloes. The most stringent combination of the two criteria ($N
    _{\rm prog}<2$ plus environmental pollution) is shown as a purple histogram. While the $N_{\rm prog}<5$ criterion does not affect the population, the requirement of a single progenitor or the environmental pollution criterion reduce the number by up to a factor of 6.
    The top panel shows the formation rate (per unit redshift) in the simulated volume, and the bottom panel shows the cumulative number of pairs.}
    \label{fig:pairs_gen_env}
\end{figure}

Having checked how the large-scale cosmological environment affects the total number of synchronised pairs formed, we next consider the role of genetic and environmental pollution. In this case, we consider our NoMini model, in which star formation occurs one free-fall time (estimated assuming $\eta_{\rm max}=0.2$) after the ACT crossing, and a synchronisation time-scale of 5~Myr, corresponding to the time when the first SNe explode. In addition, we require the halo separation to fall in the range [0.2-0.5]~kpc. Also in this case, we assume that no prior SF occurs in mini-haloes.

The results are shown in Fig.~\ref{fig:pairs_gen_env}, where we report the total number of pairs in black (34) and the genetic-pollution filtered pairs in blue (34, requiring $N_{\rm prog}<5$) and in orange (19, with the stricter requirement of $N_{\rm prog}<2$). The environmental pollution criterion alone is shown in green (resulting in 9 pairs), whereas the most stringent combination of the two criteria, giving 6 pairs, is shown in purple.
Compared to the blue histogram in Fig.~\ref{fig:visbal}, the total number of pairs in the NoMini model (shown as a black histogram in the background) is a factor of $\sim 2$ lower, even without accounting for metal pollution. This is a consequence of the different choices in the minimum and maximum delay times for synchronisation. The minimum delay time is constant at 10~Myr in \citet{visbal14b} and density-dependent ($t_{\rm ff}(\eta_{\rm gas}=0.2)$; typically slightly longer than 10~Myr) here; the maximum  delay time is also 10~Myr \citep[in][]{visbal14b} and 5~Myr in our case. In facts, this difference results in a shorter time window and a longer time separation between the ACT crossing in the two haloes of the pair, both concurring to the reduction of the number of candidates.

If we now consider the additional criteria, we observe that the blue histogram completely overlaps with the black one, demonstrating that all candidate haloes have fewer than 5 progenitors. On the other hand, only $\sim$50\% remain if we require a single progenitor. In addition to genetic pollution, environmental pollution due to expanding metal bubbles from nearby haloes appears very important, reducing the number of synchronised pairs by about a factor of 3. Another important effect of these additional criteria can be observed in the redshift distribution of the candidates, which becomes significantly skewed towards higher redshift in the most stringent case (purple histogram), i.e. almost all the pairs form at $z>16$.
 
\subsubsection{Full models}
\begin{figure}
    \centering
    \includegraphics[width=\columnwidth,trim=0 0 0 1cm,clip]{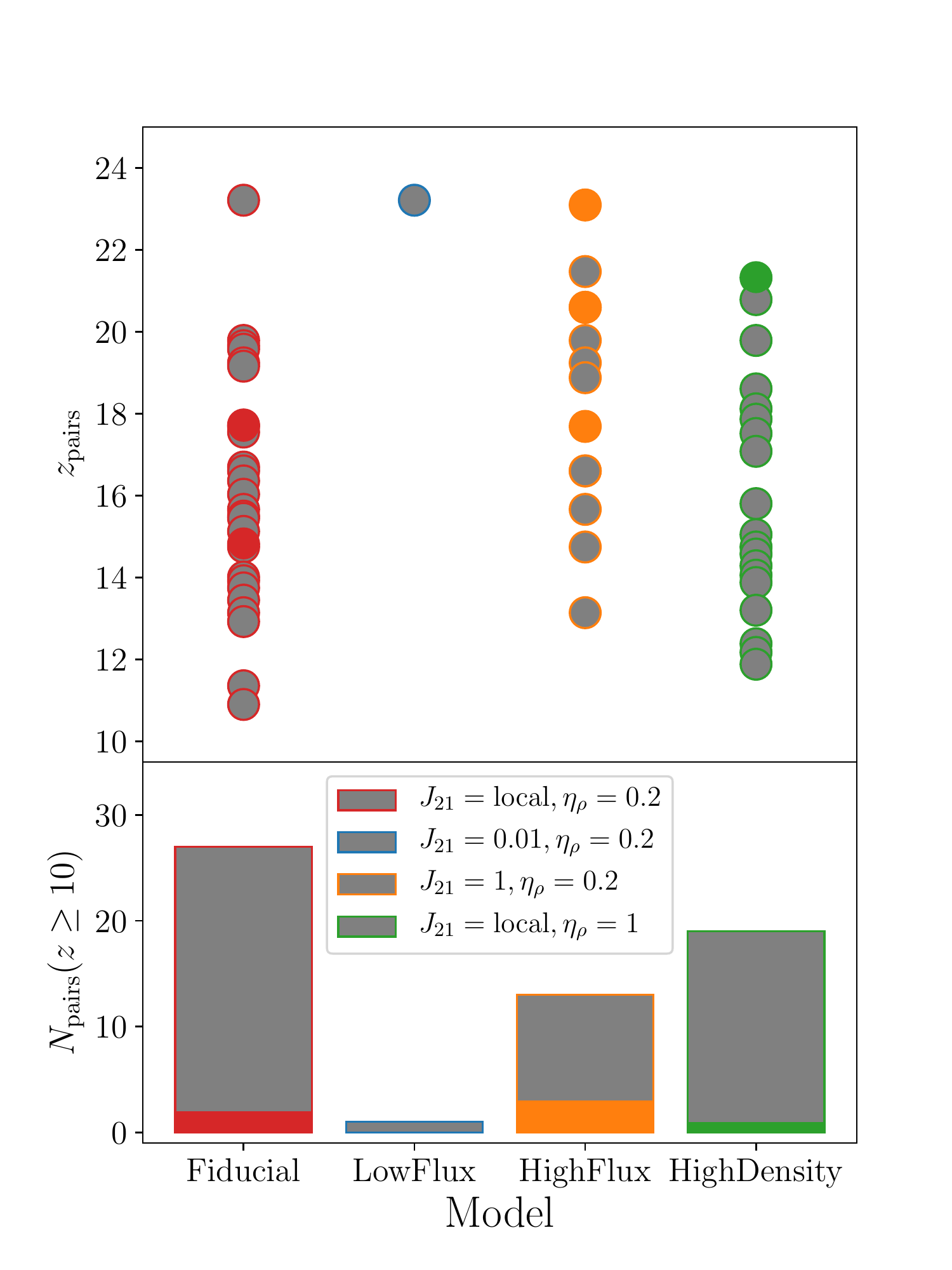}
    \caption{Comparison among the different model parameters (each column corresponding to a parameter combination, as reported in the legend). The top panel shows the redshift distribution of the synchronised pairs as grey-filled circles, whereas the coloured ones correspond to the haloes that have not been environmentally polluted. In these four models, the genetic pollution criterion is automatically verified by the lack of SF in one of the haloes of the pair.
    In the bottom panel, we show the total number of pairs down to $z=10$, with the grey histogram corresponding to the total number, and the coloured one to
    chemically pristine ones only. }
    \label{fig:model_comp}
\end{figure}

We next consider all of the physical processes described in \S~\ref{sec:method}, including SF in the mini-haloes. In this case, we have to consider two additional parameters that play a crucial role in preventing SF in low-mass haloes, i.e. the maximum density reached by the gas via adiabatic compression (defined by the $\eta$ parameter) and the LW radiation background, which sets the minimum halo mass able to efficiently form H$_2$ and cool~\citep[defined by the $J_{21}$ parameter; see, e.g.][for a discussion]{machacek01,trenti09,schauer17,kulkarni20}. In our parameter exploration, we assume $\eta=\{0.2,1.0\}$ and $J_{21}=\{0.01,1.0\}$, plus an additional case where we allow $J_{21}$ to vary locally according to Eq.~\ref{eq:j21}. An important difference from the basic models is that here, the possibility of SF to occur in mini-haloes allows us to naturally account for genetic pollution resulting from SF events in the halo progenitors, leaving environmental pollution the only additional check to be applied a posteriori.

In Fig.~\ref{fig:model_comp}, we compare our fiducial model with alternative choices of the parameters. In the top panel, we report the redshift distribution of the synchronised pair candidates, shown as grey circles, with the pristine haloes identified by the colours circles. Our fiducial model is in red, the models with constant $J_{21}$ are in blue (LowFlux; $J_{21}=0.01$) and orange (HighFlux; $J_{21}=1$) respectively, and the model with variable $J_{21}$ and $\eta_{\rho}=1$ (HighDensity) in green.
In the bottom panel, we report the cumulative number of synchronised pairs able to form down to $z=10$, with the total number in grey and the coloured histograms corresponding to the haloes that have not been environmentally polluted. While covering the redshift range $11\lsim z \lsim 20$, 
with 27 candidate pairs identified, the formation of suitable pairs in our 
fiducial model is concentrated around $z\sim 15$, with the only pristine (hence most plausible) candidates appearing at $z\sim 17.7$ and $z\sim 14.8$ (left column). If a too low $J_{21}$ is assumed, instead, almost all pairs are able to form stars (with no metal-free cases at all), because of the early formation of many stars in mini-haloes and the subsequent pollution by SNe (second column, LowFlux). This is also confirmed by the $J_{21}=1$ case (third column, HighFlux), where 13 pairs form, almost uniformly distributed in redshift, because of the constant LW radiation field that ensures SF in mini-haloes is suppressed more efficiently at the highest redshifts, thus limiting metal pollution. As a consequence, the pristine candidates in this scenario appear more easily at higher redshift relative to our fiducial case.
 
Finally, if we assume that a higher central gas density can be reached in the haloes, the locally varying $J_{21}$ becomes less effective, with 19 pairs forming,  (rightmost column, HighDensity). Moreover, only one of them is pristine, making this case slightly less favourable.

Note that all numbers above include haloes found anywhere in our simulation box. Some of the candidates we identified up to this point do not end up in the final massive quasar-host halo by $z=6$ (for example, only one of the two pristine candidates in the fiducial model does; see further discussion below).

\subsection{The dynamical heating scenario}

\begin{figure*}
    \centering
    \includegraphics[width=\columnwidth,trim=0 3cm 0 1cm,clip]{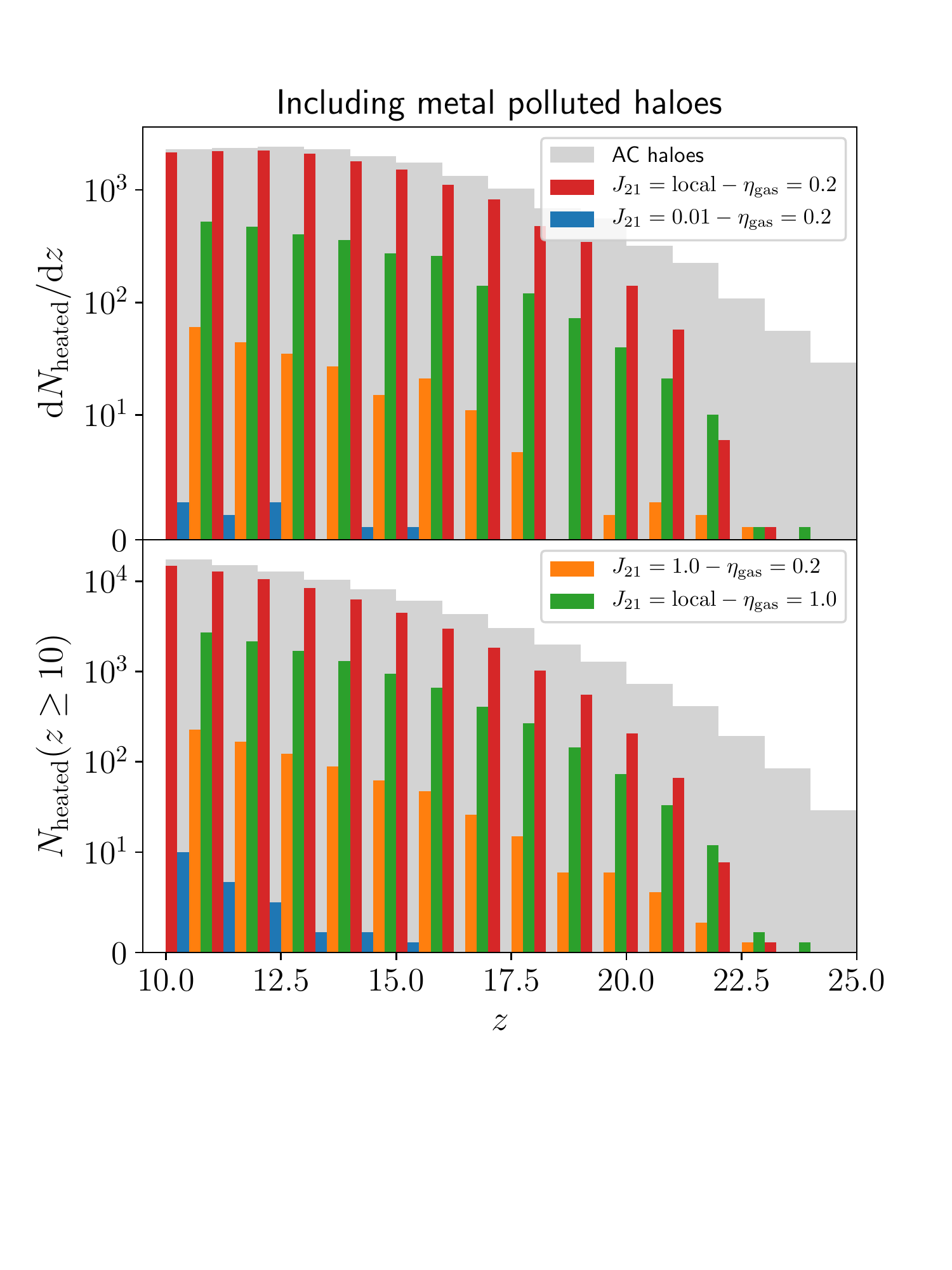}
    \includegraphics[width=\columnwidth,trim=0 3cm 0 1cm,clip]{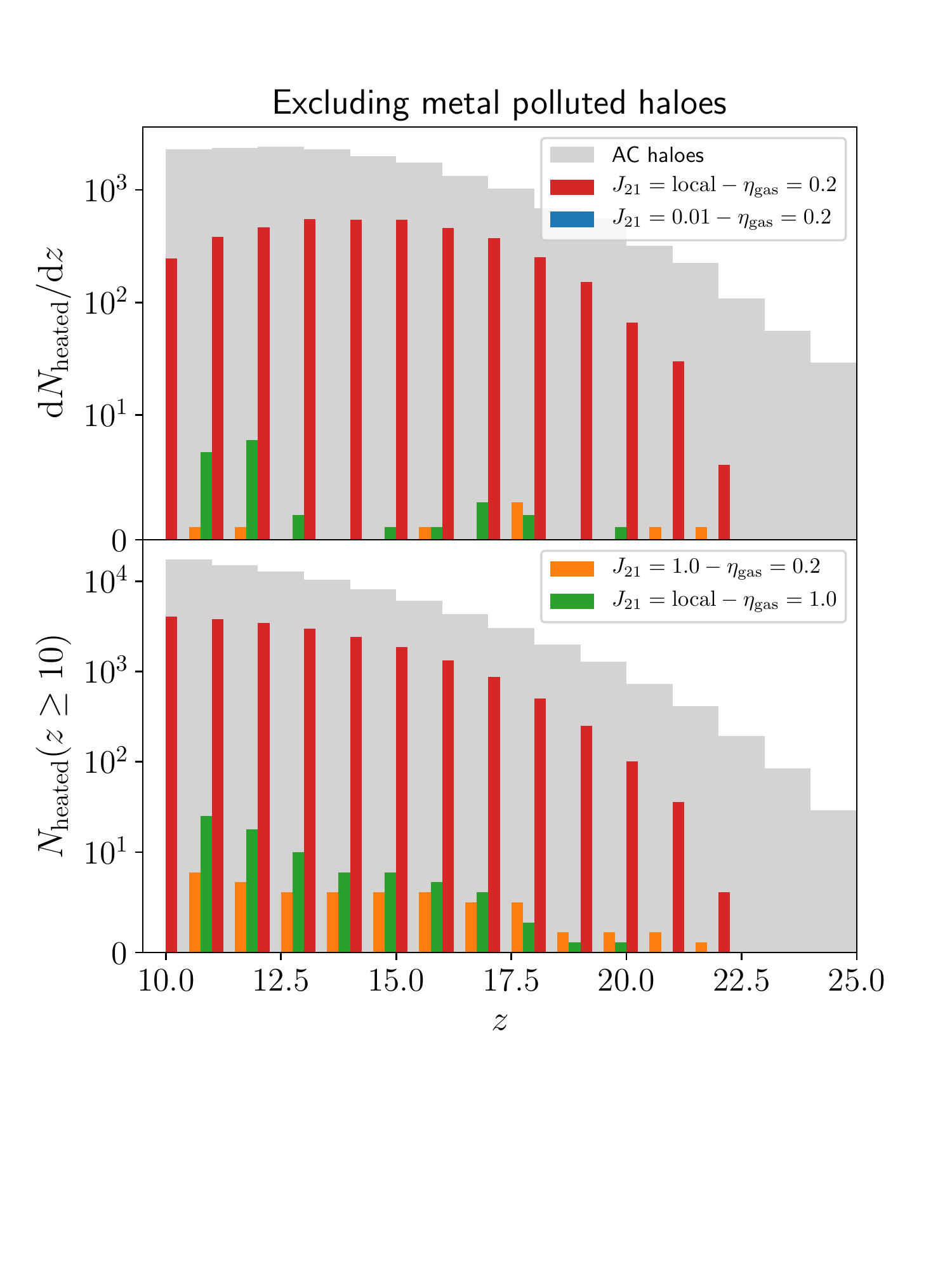}
    \caption{Redshift distribution (top panels) and cumulative number (bottom panels) of dynamically heated haloes in our model, including (left panels) or excluding (right panels) the metal-polluted haloes. The light grey histograms in the background correspond to the total population of ACHs formed in the run, and the different colours to a specific $J_{21}-\eta_{\rm gas}$ combination, as reported in the legend.}
    \label{fig:dyn_heat}
\end{figure*}

As discussed above, a possible common path to DCBH formation involves the suppression of H$_2$ cooling (and subsequent SF) in mini-haloes via dynamical heating \citep{yoshida03,inayoshi18}.  Furthermore, it has recently been suggested that these dynamically-heated halos may form very massive  stars and leave behind massive BH seeds, even though they form ${\rm H_2}$ and cool once they reach the ACH stage \citep{wise19,sakurai20} thus forming several stars that compete for accretion \citep{regan20b}, so that the seed mass is probably lower than in the synchronised pair case. Here, we consider this channel and report the results obtained in our framework, with otherwise the same set of assumptions as in the four different models discussed in the previous section. 

In Fig.~\ref{fig:dyn_heat}, we show four panels, the top ones reporting the redshift distribution of dynamically heated mini-haloes, and the bottom ones the cumulative number down to $z=10$. In the left-hand panels, we show the total number, without any constraint on the metallicity within the mini-halo. In the right-hand panels, we report the effective numbers after removal of the metal-polluted candidates. Our fiducial model is again reported in red, the LowFlux constant low-$J_{21}$ case in blue, the HighFlux constant $J_{21}=1$ case  in green, and the HighDensity variable $J_{21}$ case with $\eta_{\rho}=1$ in green. The light grey histogram in the background shows the total number of ACHs forming in the region (per redshift bin and cumulative in the top and bottom panels, respectively). 

Interestingly, if we do not consider metal pollution, our fiducial case produces 14,963 dynamically-heated haloes, i.e. about 85\% of the total population, with the redshift distribution closely matching that of the total number of ACHs. This is due to the large $J_{21}$ produced by the few star-forming mini-haloes and by the large population of ACHs (which form stars at a much higher rate than mini-haloes), and the relatively compact size of the region considered (with volume $\sim 8\rm\, cMpc^3$ at $z=6$, enclosing a total mass of $\sim 6.5\times 10^{12}\rm\, M_\odot$). The number of candidates significantly drops when we consider either a higher gas density (2,698; green histogram, HighDensity) or a constant moderately high LW background (228; orange histogram, HighFlux). In the most pessimistic case LowFlux of an extremely low LW background, instead, the number of dynamically heated haloes becomes extremely small (10; blue histogram), because of the low minimum halo mass able to cool via H$_2$ cooling and the relatively high 
dynamical heating rate needed to overcome cooling.

If we apply the metal pollution constraints, the constant $J_{21}$ cases see a net reduction in number (to about 27\% in our fiducial model and about 10\% in the other cases), which result in the lack of any dynamically heated haloes in the $J_{21}=0.01$ case, and a shift to higher redshifts of the peak of the distribution. As for the variable LW flux cases, instead, our reference case exhibits a one order of magnitude decrease in number, with the peak reached around $z\sim 15$, whereas the higher $\eta_\rho$ case is almost completely suppressed, because of the typically shorter cooling timescales in the haloes, resulting in higher SFRs and subsequent metal pollution of their neighbourhood. In general, in most of the cases explored, the number of dynamically heated haloes is larger than that of synchronised pairs, suggesting that the conditions to dynamically heat mini-haloes are easier to realise than those guaranteeing halo synchronisation, especially in the overdense environment typical of massive galaxies at high-redshift. 

\begin{figure*}
    \centering
    \includegraphics[width=\textwidth,trim=0 0cm 2cm 0,clip]{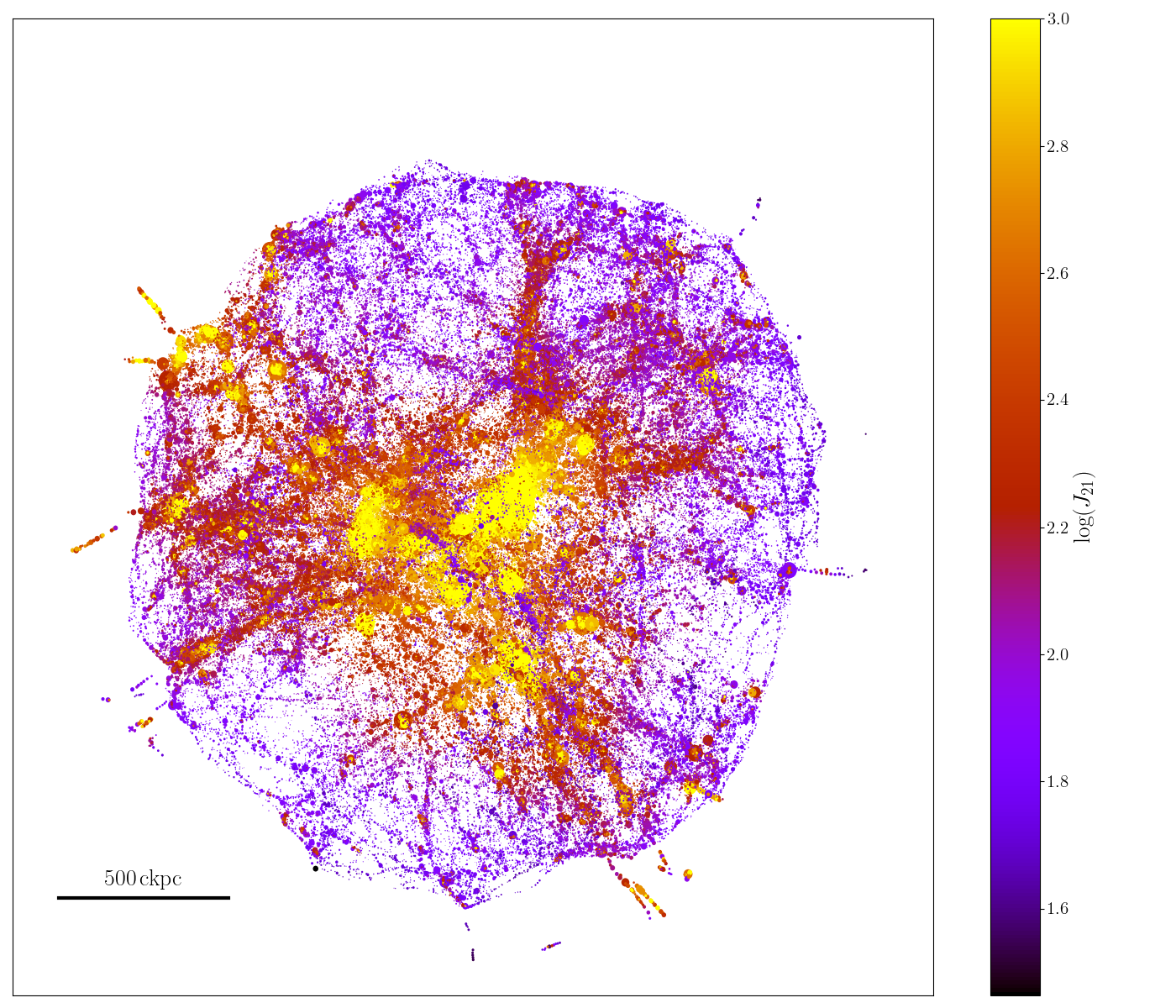}
    \caption{Projection of the $J_{21}$ distribution in our high-resolution region at $z=10$. While the low-density regions exhibit a low $J_{21}$, as expected, the most massive filaments and knots in the cosmic web, where most of the haloes form, show a significantly higher LW flux, reaching a peak of more than 500 in the centre, where the most massive haloes (progenitors of the expected quasar host), i.e. the most rapidly star-forming sites, form. Such a high LW flux does not necessarily imply a more likely formation of DCBHs in the central region, because of the correspondingly higher metal pollution due to the large number of SNe occurring in the same region.}
    \label{fig:j21map}
\end{figure*}

To assess the importance of the self-consistent $J_{21}$ estimation in our analysis, we report in  Fig.~\ref{fig:j21map} the $x-y$ projection of the $J_{21}$ field in our cosmological volume at $z=10$. From the map, we can clearly notice that, because of the large over-density in the region ($\delta = 0.1$ at $z=100$, corresponding to a 5.38$\sigma$ peak),\footnote{The overdensity $\delta$ in our simulation at $z=100$ has been derived by creating a convex hull around our high-resolution region, defining the region's volume, and then deriving its average density relative to the cosmic average density. The peak height $\nu \equiv \delta/\sigma$ is then obtained using the mass variance $\sigma=0.0186$ of an $M=6.5\times 10^{12}\,\msun$ region  assuming the \citet{planck16} cosmological parameters.}
many galaxies are forming, and $J_{21}$ reaches quite high values compared to an average density region of the Universe. In particular, the most massive haloes in the region, which form at the centre of the map, are able to produce a $J_{21}$ as high as 500 or more, although we also expect the same region to be strongly polluted by the metals produced by SNe, hence not necessarily representing the most favourable location to form DCBHs (at $z=10$).

\begin{figure}
    \centering
    \includegraphics[width=\columnwidth,trim=0 0cm 0 0,clip]{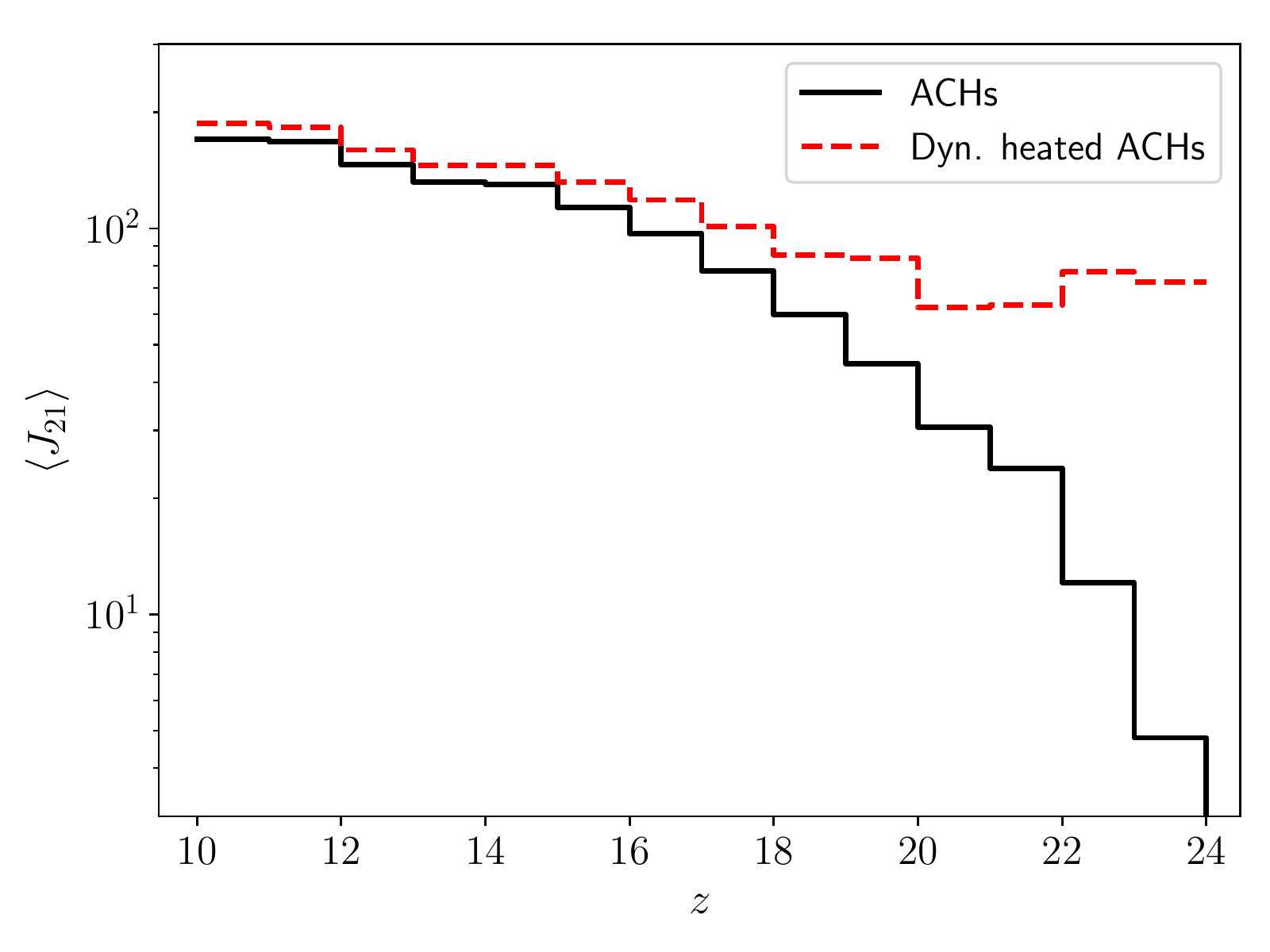}
    \caption{
    Median $J_{21}$ at the locations of the dynamically heated haloes (in red) and all the ACHs (in black) as a function of redshift for our fiducial model. The black line shows that the median $J_{21}$ in the region quickly rises (above $z=20$) and then settles around 100, a value much higher than that found in typical star forming regions at high redshift, but comparable to the values found in previous studies around massive quasar hosts \citep{valiante16}. The dynamically heated population exhibits instead a factor of 2--to--10 higher LW flux above $z=18$, consistent with their being in denser filaments around the central halo, where the number of haloes and the SFR are higher, hence producing a stronger local LW flux, able to suppress H$_2$ formation and cooling.}
    \label{fig:j21_avg}
\end{figure}

To be more quantitative, we also report in Fig.~\ref{fig:j21_avg} the median $J_{21}$ for the dynamically heated haloes in our reference model (red line), compared with the median value for the entire population of ACHs (black line), as a function of redshift. In the overdense region we consider, the value for the general population of ACHs is already higher than that of typical high-redshift star-forming regions \citep[e.g.][]{dayal18}, reaching median values of about 100 below $z=15$. While such enhancement is enough to guarantee that most of the mini-haloes are dynamically heated below $z=15$, only the haloes exposed to a 2--to-10 times higher than average LW flux can be dynamically heated at higher redshift. This can be explained with these haloes forming in denser filamentary regions (see Fig.~\ref{fig:j21map}) where the mass accretion rate is higher, and the nearby SF is able to reduce H$_2$ formation sufficiently to prevent molecular cooling. Our results are also in agreement with previous studies of high-redshift massive haloes by \citet{valiante16}, where the $J_{21}$ distribution extended to higher values than usual because of the higher SFR in the region, but also because of the larger halo clustering, an effect that cannot be tracked in semi-analytic models based on the extended Press--Schechter formalism \citep{lacey93} and its variants.

Although the high LW flux seems to suggest that dynamical heating is much more favoured compared to synchronised pairs, we have to keep in mind that as soon as the halo exceeds the ACT, H$_2$ starts to form more efficiently because of the large increase in density facilitated by atomic cooling, and a $J_{21}$ that was sufficient to prevent cooling in the mini-halo phase is not guaranteed anymore to avoid H$_2$ formation in the ACH~\citep{omukai01,oh02}.  Additionally, any metal-pollution can result in even more efficient cooling in the ACH stage. Nevertheless, the results in \citet{wise19} suggest that large accretion rates towards the centre of the halo may persist in the ACH stage, and, as mentioned above, spherically symmetric zoom-in simulations of these candidates~\citep{sakurai20} find that radiation from the growing central protostar does not shut down this rapid flow, which produces a supermassive star. \citet{regan20} also find that rapid inflow may occur even in the face of some metal pollution, but \cite{regan20b} find that widespread star formation limits the final BH mass to $<10^4 \msun$.

To summarise our results, we report in Table~\ref{tab:candidates} the number of candidate DCBH haloes in the dynamical-heating (DH; second column) and in the synchronised-pairs scenarios (SP; third column). In the last two columns we also report the number of pristine candidates for these two cases, respectively.

\begin{table}
    \centering
    \caption{Number of candidate DCBH haloes in the dynamical-heating (second column) and synchronised-pair (third column) scenarios for our models. The corresponding number of metal-free subsets of these candidates are shown in the fourth and fifth columns. Note that not all of these candidates end up in the massive quasar-host halo by $z=6$ (see Table~\ref{tab:fate} below).}
    \begin{tabular}{lcccc}
        \hline
            Model & DH & SP & Pristine DH & Pristine SP \\
        \hline
         Fiducial &  14698 & 27 & 4065 & 2\\
         LowFlux  &     10 &  1 & 0 & 0\\
         HighFlux &    228 & 13 & 8 & 3\\
         HighDensity &   2698 & 19 & 25 & 1\\
         \hline
    \end{tabular}
    
    \label{tab:candidates}
\end{table}

\subsection{The fate of the BH seed candidate haloes}

Although we have assessed the number of candidate progenitor haloes expected to form in the overdense region that subsequently evolves to a high-redshift massive halo, our analysis so far has not considered the subsequent evolution of these haloes, i.e. if they are going to merge with the main progenitor of our target halo by $z=6$ or not. Indeed, some of these haloes could simply end up as satellites, or even as nearby haloes outside the virial radius of the target halo, possibly merging with it at a lower redshift. To check if these candidates are likely sites where the quasar seed BH can form and grow to $\sim 10^9~{\rm M_\odot}$, as observed by $z\approx 6$, we checked how many synchronised pairs and dynamically-heated haloes belong to the merger history of our target halo. The results are reported in Table~\ref{tab:fate} for the same four cases discussed above, with the main results of our analysis highlighted in bold.
For both scenarios, about 30--to--70 per cent of the candidates appear as progenitors of the target halo, i.e. 9 synchronised pairs and 5,463 dynamically heated haloes in our fiducial model. 

When accounting for metal enrichment, excluding polluted haloes from the candidate list, the numbers get reduced to about one third (or less) of the total, which translates in `only' one synchronised pair and 1,389 dynamically heated haloes in our fiducial model. The large number of dynamically heated haloes can be easily explained by the strong $J_{21}$ produced by the highly star-forming galaxies in the simulated region. These galaxies create a large LW "bubble" (i.e. a region in which the local LW flux dominates the global background) around the target halo, able to more efficiently suppress H$_2$ formation in the mini-haloes around it, hence boosting the number of dynamically-heated mini-haloes in its surrounding. This does not mean that the seed BH formation via dynamical heating is less likely in the target halo (the number is typically higher than that of synchronised pairs anyway), simply that the overdensity boost affects also nearby halos. The formation of many seeds within a relatively small volume is favourable to the eventual formation of MBH pairs as these halos merge with one another during their cosmic assembly.

\begin{table}
    \centering
    \caption{Number of synchronised pairs (SP) and dynamically heated (DH) haloes belonging to the merger history of the target halo in our simulation (QSO prog) in the four models considered in the analysis (second and third columns). The number of pristine halos among these candidates is shown in the fourth and fifth columns, respectively, and is highlighted in bold face as the main result of this paper.  The number in the parentheses corresponds to the pristine fraction (percentage) of these progenitors.}
    
    \begin{tabular}{lcccc}
    \hline
    Model & \multicolumn{2}{c}{QSO prog} & \multicolumn{2}{c}{Pristine QSO prog (\%)} \\
             &        DH & SP & DH & SP \\
    \hline
     Fiducial   &  5463 &  9 & \textbf{1389} (25.4\%) & \textbf{1} (11.1\%) \\
     LowFlux    &     7 &  1 & \textbf{   0} ( 0.0\%) & \textbf{0} ( 0.0\%) \\
     HighFlux   &    90 &  8 & \textbf{   4} ( 4.4\%) & \textbf{3} (37.5\%) \\
     HighDensity&  1359 & 11 & \textbf{   7} ( 0.5\%) & \textbf{1} ( 9.1\%)\\
    \hline
    \end{tabular}
    \label{tab:fate}
\end{table}

\section{Caveats}
\label{sec:caveats}
Despite the large parameter variation we explored here, there are some caveats in our analysis that must be considered. First, the assumption of an 100\% LW radiation escape fraction can result is a more effective suppression of H$_2$ formation, hence cooling, boosting the population of both dynamically heated haloes and synchronised pairs (although the effect is more moderate for the latter case). Indeed, while in low-mass haloes gas is strongly affected by stellar feedback, likely leaving LW radiation to escape efficiently, more massive haloes could instead exhibit lower escape fractions, reducing the effective $J_{21}$ reaching the nearby mini-haloes \citep[see, e.g][]{schauer17}.

Secondly, the model we employed for metal bubble propagation, despite the improvements we made by considering the lower densities outside the haloes, is still very approximate and does not consider how the bubble expansion velocity and shape change due to mixing \citep[see][for a discussion]{2016MNRAS.463..529H}. This could result in either an overestimation or underestimation of the metal pollution, which then affects gas cooling and the possible formation of DCBHs. Moreover, the assumption of a constant SFR in Eq.~\eqref{eq:rbub}, equal to the average SFR since the first SF burst could also result in an overprediction/underprediction of the bubble size, since younger stars would create metal bubbles at later times in possibly different density conditions, and would then need time to propagate outwards.

Third, although the mass resolution in our simulation is already high, allowing us to properly resolve the ACT, it is still too low to resolve the lowest-mass minihalos, near the minimum halo mass for H$_2$ cooling for low and moderate LW fluxes. This inability might in principle result in the overprediction of the DCBH candidates (see  Section~\ref{sec:method}).

\section{Discussion}
\label{sec:discussion}

Our results highlight how a non-negligible number of synchronised pairs can form in the overdense regions where quasar hosts are expected to form, as long as the background LW radiation field is larger than $J_{21}=0.01$. Although most of these pairs would be already contaminated by metals in our models, employing simple models for metal propagation and mixing, we find that some of these haloes remain plausibly pristine DCBH host candidates. In our 
fiducial case, we find that only two pairs match all the conditions to be a plausible DCBH host candidate, of which only one belongs to the merger history of the quasar host.  If we relax the spatial separation condition for synchronised pairs from [0.25-0.5]~kpc to the intervals explored by \citet{visbal14b}, as shown in Section~\ref{sec:visbal}, the number of plausible candidates belonging to the quasar host merger history increases from 9 to 48 (up to 750~pc) and 94 (up to 1~kpc), of which 7 and 19 are pristine, respectively.  In general, our findings confirm that the large overdensity in the region where a massive quasar host later forms exhibits ideal conditions for the formation of close pairs.  Because of the semi-analytic nature of our framework, a detailed analysis of the gas evolution within the halo is not possible, and is deferred to a future study.

As for dynamical heating, our results suggest that a very large number of dynamically heated ACHs could form in the overdense regions where quasar hosts subsequently assemble (14,698 in our fiducial model, of which 1,522 appear between $z=16$ to $z=15$). More importantly, 5,463 of these ($\sim 37\%$) are progenitors of the quasar host, and 1,389 are also pristine, hence possible DCBH candidates, as long as the the accretion rate after the ACT has been crossed remains large (see \citealt{regan20} and discussion above). Our Lagrangian volume was centred around a $3\times 10^{12}\msun$ halo at $z=6$. \citet{wise19} considered a somewhat less overdense region (a Lagrangian volume  centred around an $3\times 10^{10}\msun$ halo at $z=6$) 
and found about 600 such haloes between $z=16$ and $z=15$. 
Interestingly, we also observe that dynamical heating represents a viable and relevant path to reach the ACT without forming stars as long as a moderate LW background ($J_{21}\gtrsim 1$) is present, able to simply reduce the equilibrium fraction of H$_2$ forming in the haloes, rather than completely suppressing it. This result is also in line  with \citet{wise19}, where the dynamically heated haloes exhibited $f_{\rm H_2}\sim 10^{-6}-10^{-5}$, much smaller than the values in the absence of any LW flux \citep{haiman96,tegmark97}.
On the other hand, if no LW radiation is present, the accretion rates needed to balance H$_2$ cooling would be unrealistically large. 

Overall, we note that the fulfilment of the dynamical heating criterion does not guarantee that the ACH will collapse without fragmenting.  This is because even if the heating rate matches the H$_2$ cooling in the mini-halo stage, the subsequent evolution of the ACH is more uncertain. Indeed, once the ACT is crossed, cooling by atomic H lines (and metals, if present), together with the rapid change in the thermodynamic conditions (gas becomes ionised after the ACT has been exceeded, just before atomic cooling is activated), could still boost the production of H$_2$, hence favouring fragmentation and reducing the accretion rate towards the centre, preventing the formation of a DCBH. This is in line with \citet{wise19}, who found that in the DCBH host candidates they identified in their hydrodynamic simulations, dynamical heating is insufficient to prevent H$_2$ formation during the ACH stage.  Nevertheless, in a few such haloes, they also found that the accretion rates remained high, plausibly leading to the formation of a supermassive star and a DCBH. \citet{sakurai20} recently followed the subsequent evolution of these candidates in spherically symmetric radiation-hydrodynamic simulations, and indeed found that the collapse is not prevented by radiative feedback, and produces a supermassive star, suggesting that dynamical heating alone may be a sufficient condition to lead to the formation of a DCBH. In a more realistic cosmological environment, however, \citet{regan20b} find that accretion is suppressed shortly after the formation of the first star by the formation of several other stars that compete for accretion and by the motion of the star(s) in the messy non-spherical potential of high-redshift galaxies, resulting in DCBHs of about $10^3\,\msun$. While such masses can be enough to explain billion solar mass MBHs at $z\sim 6$, they still represent a potential problem to explain MBHs above $10^8\, \msun$ at $z>7$, even assuming an optimistic sustained accretion at the Eddington limit.

Synchronised pairs, on the other hand, are fundamentally physically different. As already discussed above, cooling via H$_2$ can be avoided by some heating or H$_2$-destruction mechanisms. However, both of them depend in a roughly linear way on the density, and become $\sim 4$ orders of magnitude shorter once the ACT is crossed. This is because, once the ACT is crossed, Ly$\alpha$ cooling allows the central density in the halo to increase rapidly. It is therefore impossible for dynamical heating to keep up with H$_2$ formation and cooling beyond this point. On the other hand, the LW flux (from a nearby star-forming ACH) can still be large enough to continue suppressing H$_2$ formation even at these higher densities, up to $n\sim 10^4\rm\, cm^{-3}$, which corresponds to the critical density of H$_2$, above which H$_2$ cooling becomes ineffective. This is due to collisional dissociation disabling H$_2$ formation, and the H$_2$ populations saturating near \textit{local thermodynamic equilibrium} levels, that inhibit H$_2$ cooling, as explicitly demonstrated by \citet{fernandez14}. Therefore, while dynamical heating can never keep H$_2$ from
cooling the gas in ACHs, a sufficiently large LW flux could in principle completely stop it, and suppress fragmentation. Several papers \citep{shang10,regan14,fernandez14,becerra15} have indeed studied the behaviour of gas collapsing in H$_2$-suppressed ACHs, at increasingly high
resolution.  While some modest amounts of fragmentation were nonetheless found
in some cases, only a handful of clumps were able to form. These clumps
still accrete at the high rates required for supermassive star formation, and,
additionally, most of these clumps collide and merge quickly before
they could evolve into main-sequence stars \citep{inayoshi20}.

Therefore, given the possible limitations on the accretion rate in dynamically heated mini-haloes, pair synchronisation still appears to be the best case to trigger massive DCBH formation.
While we find a large number of mini-haloes that are dynamically heated and avoid H$_2$ cooling, the vast majority of these fail the synchronised-pair criterion (we find 9 synchronised pairs that end up in the quasar-host halo at $z=6$ but only one of these is metal-free).

\begin{figure}
    \centering
    \includegraphics[width=\columnwidth]{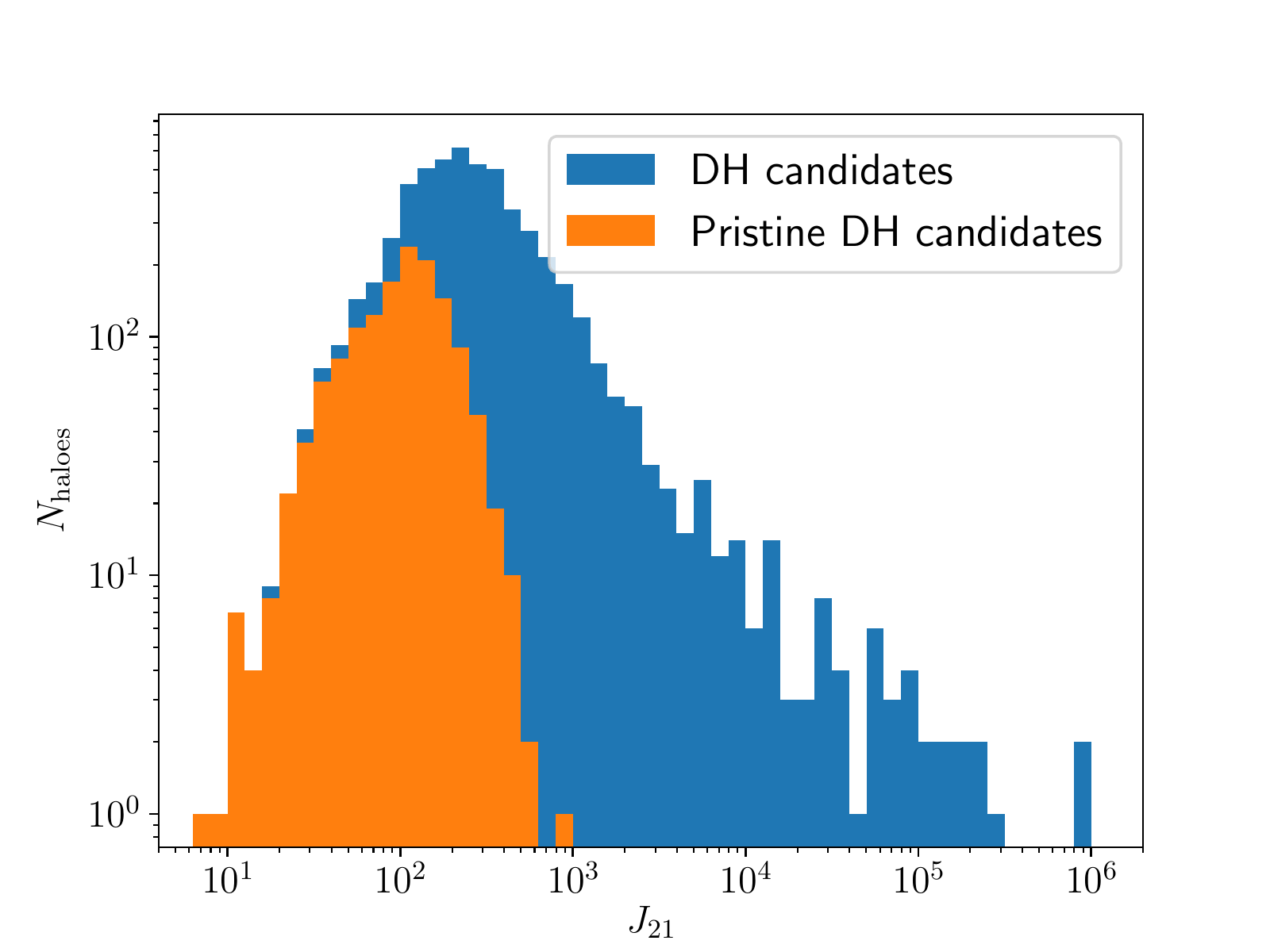}
    \caption{$J_{21}$ distribution in dynamically heated mini-haloes (at the time of the ACT crossing) from our fiducial model, including (blue histogram) or excluding (orange histogram) metal polluted haloes.}
    \label{fig:j21dist}
\end{figure}

However, we note that the synchronisation criterion, requiring a close pair of ACHs, is not the only way to guarantee the high LW flux necessary for DCBH formation. It is possible to obtain such fluxes through the combined effect of multiple nearby LW sources, a scenario that we can name \textit{synchronised multiplets}. To test whether these conditions actually occur in our overdense region, we evaluated the $J_{21}$ distribution sampled by dynamically heated mini-haloes in our fiducial model, excluding haloes already flagged as synchronised pairs. The results are shown in Fig.~\ref{fig:j21dist}, where the blue histogram corresponds to the full distribution (including metal polluted haloes), and the orange one to the pristine haloes only. Although the full distribution extends up to $J_{21}\sim 10^6$, most of the haloes in the high-LW flux tail are metal-rich, as such high $J_{21}$ values can be reached only near the very massive (hence highly star-forming) haloes present in the region \citep[see also][]{habouzit16}. When we remove the metal polluted haloes, the distribution shifts to lower LW fluxes, but with 38 haloes still exhibiting $J_{21}>300$, and one of them even reaching $\sim 850$. Even though 300 is on the low end of the range of critical $J_{21}$ values considered for efficient DCBH formation \citep{dijkstra14}, the non-negligible number of such haloes, and most importantly the fact that one of them also exhibits a value close to a more conservative $J_{21}=10^3$, hint that the clustering of LW sources  
in overdense regions might represent a viable alternative for DCBH formation to pair synchronisation.

\section{Conclusions}
\label{sec:conclusions}

In this work, we combined a semi-analytic framework with N-body simulations, in order to assess the formation of DCBHs in overdense regions where high-redshift quasar hosts are expected to form. The framework exploited the dark-matter-only version of the simulation presented in \citet{lupi19b}, where the particle mass resolution is high enough too resolve the last stages of the mini-halo phase and the transition to the atomic cooling halo phase.

We considered two different mechanisms proposed to be implicated in  DCBH formation, the "synchronised pairs" scenario \citep{dijkstra08,agarwal12,dijkstra14,visbal14b,habouzit17,chon18,regan19}, where the time and spatial proximity between two ACHs (of which one has already formed stars) guarantees a strong enough LW flux to suppress H$_2$ formation \citep{shang10,wolcott-green11,latif14,wolcott-green19}, and the "dynamical heating" scenario \citep{yoshida03,inayoshi18,wise19}, where  H$_2$ formation is suppressed by compressional heating caused by rapid halo growth.

The goal of this paper is to identify halos that meet  the following criteria:  \textit{(i)} metal-free, \textit{(ii)} belonging to the quasar host merger history,  \textit{(iii)} star formation disabled
in the mini-halo phase (because of the combined effect of dynamical heating and a moderate  LW flux), and \textit{(iv)}  H$_2$ formation and cooling significantly suppressed in the atomic-cooling phase, because of the presence of a nearby star-forming ACH (a synchronised pair).  While satisfying the first three conditions is expected to produce unusually massive stars ($\sim 10^{3-4}~{\rm M_\odot}$), all four conditions are likely necessary to produce supermassive stars leaving BH remnants as massive as $10^{5-6}~{\rm M_\odot}$.

This work shows that the very dense environments that lead to the formation of the massive quasar hosts at high redshift lead to a high number of ACHs in a relatively small volume.  A large fraction (as high as 85\%) of these is dynamically heated during their mini-halo phase, helped by the high local levels of LW radiation due to a strong clustering of nearby sources.
The number density of synchronised pairs is also higher, by approximately an order of magnitude, than in average density regions of the Universe. In particular,  the synchronised pairs identified have avoided SF during the mini-halo phase thanks to the combined effect of dynamical heating and the higher LW flux in the region.

Metal pollution decreases by a factor of about five the number of halos eligible to host DCBH formation. In our fiducial case we found one pristine synchronised pair of halos and $\sim 1400$ pristine dynamically heated halos. The location of the candidates also is an important factor: we find $\sim 4000$ dynamically heated pristine halos and 2 synchronised pairs in the simulation volume, but only about half of them belong to halos in the main halo merger history, therefore able to seed the quasar. The number of candidates increases up to a few tens if we consider separations up to 1~kpc in the synchronised pairs or halo multiplets in which the  overlapping LW radiation from different star forming halos boosts the total impinging flux.

Our results here thus suggest that at least one massive seed from a synchronised pair and several lower-mass seeds from dynamically heated halos could form in this region, and grow and merge with other similar massive BH seeds during the cosmic history. This conclusion is important for gravitational wave observatories such as the Laser Interferometer Space Antenna (\textit{LISA}) and for third-generation terrestrial instruments, which will be able to detect mergers between seed BHs up to the redshift of their formation with high signal-to-noise.

\section*{Acknowledgements}
We thank the anonymous referee for their useful comments.
AL acknowledges support from MIUR under the grant PRIN 2017-MB8AEZ. 
ZH acknowledges support from NSF grants 2006176 and 1715661, and NASA grant NNX17AL82G.
The analysis reported in this work has been performed on the Horizon cluster, hosted by the Institut d’Astrophysique de Paris.

\section*{Data Availability Statement}
The data underlying this article will be shared on reasonable request to the corresponding author.



\bibliographystyle{mnras}
\bibliography{Biblio} 

\begin{thebibliography}{}
\makeatletter
\relax
\def\mn@urlcharsother{\let\do\@makeother \do\$\do\&\do\#\do\^\do\_\do\%\do\~}
\def\mn@doi{\begingroup\mn@urlcharsother \@ifnextchar [ {\mn@doi@}
  {\mn@doi@[]}}
\def\mn@doi@[#1]#2{\def\@tempa{#1}\ifx\@tempa\@empty \href
  {http://dx.doi.org/#2} {doi:#2}\else \href {http://dx.doi.org/#2} {#1}\fi
  \endgroup}
\def\mn@eprint#1#2{\mn@eprint@#1:#2::\@nil}
\def\mn@eprint@arXiv#1{\href {http://arxiv.org/abs/#1} {{\tt arXiv:#1}}}
\def\mn@eprint@dblp#1{\href {http://dblp.uni-trier.de/rec/bibtex/#1.xml}
  {dblp:#1}}
\def\mn@eprint@#1:#2:#3:#4\@nil{\def\@tempa {#1}\def\@tempb {#2}\def\@tempc
  {#3}\ifx \@tempc \@empty \let \@tempc \@tempb \let \@tempb \@tempa \fi \ifx
  \@tempb \@empty \def\@tempb {arXiv}\fi \@ifundefined
  {mn@eprint@\@tempb}{\@tempb:\@tempc}{\expandafter \expandafter \csname
  mn@eprint@\@tempb\endcsname \expandafter{\@tempc}}}

\bibitem[\protect\citeauthoryear{{Abel}, {Bryan}  \& {Norman}}{{Abel}
  et~al.}{2002}]{abel02}
{Abel} T.,  {Bryan} G.~L.,   {Norman} M.~L.,  2002, \mn@doi [Science]
  {10.1126/science.295.5552.93}, \href
  {http://adsabs.harvard.edu/abs/2002Sci...295...93A} {295, 93}

\bibitem[\protect\citeauthoryear{{Agarwal}, {Khochfar}, {Johnson}, {Neistein},
  {Dalla Vecchia}  \& {Livio}}{{Agarwal} et~al.}{2012}]{agarwal12}
{Agarwal} B.,  {Khochfar} S.,  {Johnson} J.~L.,  {Neistein} E.,  {Dalla
  Vecchia} C.,   {Livio} M.,  2012, \mn@doi [\mnras]
  {10.1111/j.1365-2966.2012.21651.x}, \href
  {http://adsabs.harvard.edu/abs/2012MNRAS.425.2854A} {425, 2854}

\bibitem[\protect\citeauthoryear{{Alexander} \& {Natarajan}}{{Alexander} \&
  {Natarajan}}{2014}]{alexander14}
{Alexander} T.,  {Natarajan} P.,  2014, \mn@doi [Science]
  {10.1126/science.1251053}, \href
  {https://ui.adsabs.harvard.edu/abs/2014Sci...345.1330A} {345, 1330}

\bibitem[\protect\citeauthoryear{{Ba{\~n}ados} et~al.,}{{Ba{\~n}ados}
  et~al.}{2018}]{banados18}
{Ba{\~n}ados} E.,  et~al., 2018, \mn@doi [\nat] {10.1038/nature25180}, \href
  {http://adsabs.harvard.edu/abs/2018Natur.553..473B} {553, 473}

\bibitem[\protect\citeauthoryear{{Balberg}, {Shapiro}  \& {Inagaki}}{{Balberg}
  et~al.}{2002}]{balberg02}
{Balberg} S.,  {Shapiro} S.~L.,   {Inagaki} S.,  2002, \mn@doi [\apj]
  {10.1086/339038}, \href
  {https://ui.adsabs.harvard.edu/abs/2002ApJ...568..475B} {568, 475}

\bibitem[\protect\citeauthoryear{{Barkana}}{{Barkana}}{2004}]{barkana04}
{Barkana} R.,  2004, \mn@doi [\mnras] {10.1111/j.1365-2966.2004.07177.x}, \href
  {https://ui.adsabs.harvard.edu/abs/2004MNRAS.347...59B} {347, 59}

\bibitem[\protect\citeauthoryear{{Barkana} \& {Loeb}}{{Barkana} \&
  {Loeb}}{2001}]{barkana01}
{Barkana} R.,  {Loeb} A.,  2001, \mn@doi [\physrep]
  {10.1016/S0370-1573(01)00019-9}, \href
  {http://adsabs.harvard.edu/abs/2001PhR...349..125B} {349, 125}

\bibitem[\protect\citeauthoryear{{Becerra}, {Greif}, {Springel}  \&
  {Hernquist}}{{Becerra} et~al.}{2015}]{becerra15}
{Becerra} F.,  {Greif} T.~H.,  {Springel} V.,   {Hernquist} L.~E.,  2015,
  \mn@doi [\mnras] {10.1093/mnras/stu2284}, \href
  {https://ui.adsabs.harvard.edu/abs/2015MNRAS.446.2380B} {446, 2380}

\bibitem[\protect\citeauthoryear{{Begelman}}{{Begelman}}{1979}]{begelman79}
{Begelman} M.~C.,  1979, \mn@doi [\mnras] {10.1093/mnras/187.2.237}, \href
  {https://ui.adsabs.harvard.edu/abs/1979MNRAS.187..237B} {187, 237}

\bibitem[\protect\citeauthoryear{{Begelman}}{{Begelman}}{2010}]{begelman10}
{Begelman} M.~C.,  2010, \mn@doi [\mnras] {10.1111/j.1365-2966.2009.15916.x},
  \href {http://adsabs.harvard.edu/abs/2010MNRAS.402..673B} {402, 673}

\bibitem[\protect\citeauthoryear{{Begelman}, {Volonteri}  \& {Rees}}{{Begelman}
  et~al.}{2006}]{begelman06}
{Begelman} M.~C.,  {Volonteri} M.,   {Rees} M.~J.,  2006, \mn@doi [\mnras]
  {10.1111/j.1365-2966.2006.10467.x}, \href
  {http://adsabs.harvard.edu/abs/2006MNRAS.370..289B} {370, 289}

\bibitem[\protect\citeauthoryear{{Begelman}, {Rossi}  \& {Armitage}}{{Begelman}
  et~al.}{2008}]{begelman08}
{Begelman} M.~C.,  {Rossi} E.~M.,   {Armitage} P.~J.,  2008, \mn@doi [\mnras]
  {10.1111/j.1365-2966.2008.13344.x}, \href
  {http://cdsads.u-strasbg.fr/abs/2008MNRAS.387.1649B} {387, 1649}

\bibitem[\protect\citeauthoryear{{Behroozi}, {Wechsler}  \& {Wu}}{{Behroozi}
  et~al.}{2013a}]{behroozi13}
{Behroozi} P.~S.,  {Wechsler} R.~H.,   {Wu} H.-Y.,  2013a, \mn@doi [\apj]
  {10.1088/0004-637X/762/2/109}, \href
  {http://adsabs.harvard.edu/abs/2013ApJ...762..109B} {762, 109}

\bibitem[\protect\citeauthoryear{{Behroozi}, {Wechsler}, {Wu}, {Busha},
  {Klypin}  \& {Primack}}{{Behroozi} et~al.}{2013b}]{behroozi11}
{Behroozi} P.~S.,  {Wechsler} R.~H.,  {Wu} H.-Y.,  {Busha} M.~T.,  {Klypin}
  A.~A.,   {Primack} J.~R.,  2013b, \mn@doi [\apj]
  {10.1088/0004-637X/763/1/18}, \href
  {https://ui.adsabs.harvard.edu/abs/2013ApJ...763...18B} {763, 18}

\bibitem[\protect\citeauthoryear{{Boekholt}, {Schleicher}, {Fellhauer},
  {Klessen}, {Reinoso}, {Stutz}  \& {Haemmerl{\'e}}}{{Boekholt}
  et~al.}{2018}]{boekholt18}
{Boekholt} T.~C.~N.,  {Schleicher} D.~R.~G.,  {Fellhauer} M.,  {Klessen} R.~S.,
   {Reinoso} B.,  {Stutz} A.~M.,   {Haemmerl{\'e}} L.,  2018, \mn@doi [\mnras]
  {10.1093/mnras/sty208}, \href
  {https://ui.adsabs.harvard.edu/abs/2018MNRAS.476..366B} {476, 366}

\bibitem[\protect\citeauthoryear{{Choi}, {Shlosman}  \& {Begelman}}{{Choi}
  et~al.}{2013}]{choi13}
{Choi} J.-H.,  {Shlosman} I.,   {Begelman} M.~C.,  2013, \mn@doi [\apj]
  {10.1088/0004-637X/774/2/149}, \href
  {http://adsabs.harvard.edu/abs/2013ApJ...774..149C} {774, 149}

\bibitem[\protect\citeauthoryear{{Chon}, {Hirano}, {Hosokawa}  \&
  {Yoshida}}{{Chon} et~al.}{2016}]{chon16}
{Chon} S.,  {Hirano} S.,  {Hosokawa} T.,   {Yoshida} N.,  2016, \mn@doi [\apj]
  {10.3847/0004-637X/832/2/134}, \href
  {https://ui.adsabs.harvard.edu/abs/2016ApJ...832..134C} {832, 134}

\bibitem[\protect\citeauthoryear{{Chon}, {Hosokawa}  \& {Yoshida}}{{Chon}
  et~al.}{2018}]{chon18}
{Chon} S.,  {Hosokawa} T.,   {Yoshida} N.,  2018, \mn@doi [\mnras]
  {10.1093/mnras/sty086}, \href
  {https://ui.adsabs.harvard.edu/abs/2018MNRAS.475.4104C} {475, 4104}

\bibitem[\protect\citeauthoryear{{D'Amico}, {Panci}, {Lupi}, {Bovino}  \&
  {Silk}}{{D'Amico} et~al.}{2018}]{damico18}
{D'Amico} G.,  {Panci} P.,  {Lupi} A.,  {Bovino} S.,   {Silk} J.,  2018,
  \mn@doi [\mnras] {10.1093/mnras/stx2419}, \href
  {https://ui.adsabs.harvard.edu/abs/2018MNRAS.473..328D} {473, 328}

\bibitem[\protect\citeauthoryear{{Das}, {Schleicher}, {Leigh}  \&
  {Boekholt}}{{Das} et~al.}{2020}]{das20}
{Das} A.,  {Schleicher} D. R.~G.,  {Leigh} N. W.~C.,   {Boekholt} T. C.~N.,
  2020, arXiv e-prints, \href
  {https://ui.adsabs.harvard.edu/abs/2020arXiv201201456D} {p. arXiv:2012.01456}

\bibitem[\protect\citeauthoryear{{Davies}, {Miller}  \& {Bellovary}}{{Davies}
  et~al.}{2011}]{davies11}
{Davies} M.~B.,  {Miller} M.~C.,   {Bellovary} J.~M.,  2011, \mn@doi [\apjl]
  {10.1088/2041-8205/740/2/L42}, \href
  {http://adsabs.harvard.edu/abs/2011ApJ...740L..42D} {740, L42}

\bibitem[\protect\citeauthoryear{{Dayal} \& {Ferrara}}{{Dayal} \&
  {Ferrara}}{2018}]{dayal18}
{Dayal} P.,  {Ferrara} A.,  2018, \mn@doi [\physrep]
  {10.1016/j.physrep.2018.10.002}, \href
  {https://ui.adsabs.harvard.edu/abs/2018PhR...780....1D} {780, 1}

\bibitem[\protect\citeauthoryear{{Decarli} et~al.,}{{Decarli}
  et~al.}{2018}]{decarli18}
{Decarli} R.,  et~al., 2018, \mn@doi [\apj] {10.3847/1538-4357/aaa5aa}, \href
  {http://adsabs.harvard.edu/abs/2018ApJ...854...97D} {854, 97}

\bibitem[\protect\citeauthoryear{{Devecchi} \& {Volonteri}}{{Devecchi} \&
  {Volonteri}}{2009}]{devecchi09}
{Devecchi} B.,  {Volonteri} M.,  2009, \mn@doi [\apj]
  {10.1088/0004-637X/694/1/302}, \href
  {http://cdsads.u-strasbg.fr/abs/2009ApJ...694..302D} {694, 302}

\bibitem[\protect\citeauthoryear{{Devecchi}, {Volonteri}, {Colpi}  \&
  {Haardt}}{{Devecchi} et~al.}{2010}]{devecchi10}
{Devecchi} B.,  {Volonteri} M.,  {Colpi} M.,   {Haardt} F.,  2010, \mn@doi
  [\mnras] {10.1111/j.1365-2966.2010.17363.x}, \href
  {http://cdsads.u-strasbg.fr/abs/2010MNRAS.409.1057D} {409, 1057}

\bibitem[\protect\citeauthoryear{{Devecchi}, {Volonteri}, {Rossi}, {Colpi}  \&
  {Portegies Zwart}}{{Devecchi} et~al.}{2012}]{devecchi12}
{Devecchi} B.,  {Volonteri} M.,  {Rossi} E.~M.,  {Colpi} M.,   {Portegies
  Zwart} S.,  2012, \mn@doi [\mnras] {10.1111/j.1365-2966.2012.20406.x}, \href
  {http://cdsads.u-strasbg.fr/abs/2012MNRAS.421.1465D} {421, 1465}

\bibitem[\protect\citeauthoryear{{Diemer}}{{Diemer}}{2018}]{diemer18}
{Diemer} B.,  2018, \mn@doi [\apjs] {10.3847/1538-4365/aaee8c}, \href
  {https://ui.adsabs.harvard.edu/abs/2018ApJS..239...35D} {239, 35}

\bibitem[\protect\citeauthoryear{{Dijkstra}, {Haiman}, {Mesinger}  \&
  {Wyithe}}{{Dijkstra} et~al.}{2008}]{dijkstra08}
{Dijkstra} M.,  {Haiman} Z.,  {Mesinger} A.,   {Wyithe} J.~S.~B.,  2008,
  \mn@doi [\mnras] {10.1111/j.1365-2966.2008.14031.x}, \href
  {http://adsabs.harvard.edu/abs/2008MNRAS.391.1961D} {391, 1961}

\bibitem[\protect\citeauthoryear{{Dijkstra}, {Ferrara}  \&
  {Mesinger}}{{Dijkstra} et~al.}{2014}]{dijkstra14}
{Dijkstra} M.,  {Ferrara} A.,   {Mesinger} A.,  2014, \mn@doi [\mnras]
  {10.1093/mnras/stu1007}, \href
  {https://ui.adsabs.harvard.edu/abs/2014MNRAS.442.2036D} {442, 2036}

\bibitem[\protect\citeauthoryear{{Dotan}, {Rossi}  \& {Shaviv}}{{Dotan}
  et~al.}{2011}]{dotan11}
{Dotan} C.,  {Rossi} E.~M.,   {Shaviv} N.~J.,  2011, \mn@doi [\mnras]
  {10.1111/j.1365-2966.2011.19461.x}, \href
  {http://adsabs.harvard.edu/abs/2011MNRAS.417.3035D} {417, 3035}

\bibitem[\protect\citeauthoryear{{Fan} et~al.,}{{Fan} et~al.}{2006}]{fan06}
{Fan} X.,  et~al., 2006, \mn@doi [\aj] {10.1086/500296}, \href
  {http://adsabs.harvard.edu/abs/2006AJ....131.1203F} {131, 1203}

\bibitem[\protect\citeauthoryear{{Fernandez}, {Bryan}, {Haiman}  \&
  {Li}}{{Fernandez} et~al.}{2014}]{fernandez14}
{Fernandez} R.,  {Bryan} G.~L.,  {Haiman} Z.,   {Li} M.,  2014, \mn@doi
  [\mnras] {10.1093/mnras/stu230}, \href
  {https://ui.adsabs.harvard.edu/abs/2014MNRAS.439.3798F} {439, 3798}

\bibitem[\protect\citeauthoryear{{Fiacconi} \& {Rossi}}{{Fiacconi} \&
  {Rossi}}{2017}]{fiacconi17b}
{Fiacconi} D.,  {Rossi} E.~M.,  2017, \mn@doi [\mnras] {10.1093/mnras/stw2505},
  \href {https://ui.adsabs.harvard.edu/abs/2017MNRAS.464.2259F} {464, 2259}

\bibitem[\protect\citeauthoryear{{G{\"u}rkan}, {Freitag}  \&
  {Rasio}}{{G{\"u}rkan} et~al.}{2004}]{gurkan04}
{G{\"u}rkan} M.~A.,  {Freitag} M.,   {Rasio} F.~A.,  2004, \mn@doi [\apj]
  {10.1086/381968}, \href {http://cdsads.u-strasbg.fr/abs/2004ApJ...604..632G}
  {604, 632}

\bibitem[\protect\citeauthoryear{{Habouzit} et~al.,}{{Habouzit}
  et~al.}{2016a}]{habouzit16}
{Habouzit} M.,  et~al., 2016a, \mn@doi [\mnras] {10.1093/mnras/stv2740}, \href
  {https://ui.adsabs.harvard.edu/abs/2016MNRAS.456.1901H} {456, 1901}

\bibitem[\protect\citeauthoryear{{Habouzit}, {Volonteri}, {Latif}, {Dubois}  \&
  {Peirani}}{{Habouzit} et~al.}{2016b}]{2016MNRAS.463..529H}
{Habouzit} M.,  {Volonteri} M.,  {Latif} M.,  {Dubois} Y.,   {Peirani} S.,
  2016b, \mn@doi [\mnras] {10.1093/mnras/stw1924}, \href
  {https://ui.adsabs.harvard.edu/abs/2016MNRAS.463..529H} {463, 529}

\bibitem[\protect\citeauthoryear{{Habouzit}, {Volonteri}  \&
  {Dubois}}{{Habouzit} et~al.}{2017}]{habouzit17}
{Habouzit} M.,  {Volonteri} M.,   {Dubois} Y.,  2017, \mn@doi [\mnras]
  {10.1093/mnras/stx666}, \href
  {http://adsabs.harvard.edu/abs/2017MNRAS.468.3935H} {468, 3935}

\bibitem[\protect\citeauthoryear{{Haemmerl{\'e}}, {Woods}, {Klessen}, {Heger}
  \& {Whalen}}{{Haemmerl{\'e}} et~al.}{2018}]{haemmerle18}
{Haemmerl{\'e}} L.,  {Woods} T.~E.,  {Klessen} R.~S.,  {Heger} A.,   {Whalen}
  D.~J.,  2018, \mn@doi [\mnras] {10.1093/mnras/stx2919}, \href
  {https://ui.adsabs.harvard.edu/abs/2018MNRAS.474.2757H} {474, 2757}

\bibitem[\protect\citeauthoryear{{Haiman} \& {Loeb}}{{Haiman} \&
  {Loeb}}{2001}]{haiman01}
{Haiman} Z.,  {Loeb} A.,  2001, \mn@doi [\apj] {10.1086/320586}, \href
  {http://adsabs.harvard.edu/abs/2001ApJ...552..459H} {552, 459}

\bibitem[\protect\citeauthoryear{{Haiman}, {Thoul}  \& {Loeb}}{{Haiman}
  et~al.}{1996}]{haiman96}
{Haiman} Z.,  {Thoul} A.~A.,   {Loeb} A.,  1996, \mn@doi [\apj]
  {10.1086/177343}, \href {http://adsabs.harvard.edu/abs/1996ApJ...464..523H}
  {464, 523}

\bibitem[\protect\citeauthoryear{{Heger}, {Fryer}, {Woosley}, {Langer}  \&
  {Hartmann}}{{Heger} et~al.}{2003}]{heger03}
{Heger} A.,  {Fryer} C.~L.,  {Woosley} S.~E.,  {Langer} N.,   {Hartmann} D.~H.,
   2003, \mn@doi [\apj] {10.1086/375341}, \href
  {http://cdsads.u-strasbg.fr/abs/2003ApJ...591..288H} {591, 288}

\bibitem[\protect\citeauthoryear{{Hirano} \& {Bromm}}{{Hirano} \&
  {Bromm}}{2017}]{hirano17}
{Hirano} S.,  {Bromm} V.,  2017, \mn@doi [\mnras] {10.1093/mnras/stx1220},
  \href {https://ui.adsabs.harvard.edu/abs/2017MNRAS.470..898H} {470, 898}

\bibitem[\protect\citeauthoryear{{Hirano}, {Hosokawa}, {Yoshida}, {Omukai}  \&
  {Yorke}}{{Hirano} et~al.}{2015}]{hirano15}
{Hirano} S.,  {Hosokawa} T.,  {Yoshida} N.,  {Omukai} K.,   {Yorke} H.~W.,
  2015, \mn@doi [\mnras] {10.1093/mnras/stv044}, \href
  {https://ui.adsabs.harvard.edu/abs/2015MNRAS.448..568H} {448, 568}

\bibitem[\protect\citeauthoryear{{Hirano}, {Yoshida}, {Sakurai}  \&
  {Fujii}}{{Hirano} et~al.}{2018}]{hirano18}
{Hirano} S.,  {Yoshida} N.,  {Sakurai} Y.,   {Fujii} M.~S.,  2018, \mn@doi
  [\apj] {10.3847/1538-4357/aaaaba}, \href
  {https://ui.adsabs.harvard.edu/abs/2018ApJ...855...17H} {855, 17}

\bibitem[\protect\citeauthoryear{{Hopkins}}{{Hopkins}}{2015}]{hopkins15}
{Hopkins} P.~F.,  2015, \mn@doi [\mnras] {10.1093/mnras/stv195}, \href
  {http://adsabs.harvard.edu/abs/2015MNRAS.450...53H} {450, 53}

\bibitem[\protect\citeauthoryear{{Hosokawa}, {Omukai}  \& {Yorke}}{{Hosokawa}
  et~al.}{2012}]{hosokawa12}
{Hosokawa} T.,  {Omukai} K.,   {Yorke} H.~W.,  2012, \mn@doi [\apj]
  {10.1088/0004-637X/756/1/93}, \href
  {http://adsabs.harvard.edu/abs/2012ApJ...756...93H} {756, 93}

\bibitem[\protect\citeauthoryear{{Hosokawa}, {Yorke}, {Inayoshi}, {Omukai}  \&
  {Yoshida}}{{Hosokawa} et~al.}{2013}]{hosokawa13}
{Hosokawa} T.,  {Yorke} H.~W.,  {Inayoshi} K.,  {Omukai} K.,   {Yoshida} N.,
  2013, \mn@doi [\apj] {10.1088/0004-637X/778/2/178}, \href
  {http://adsabs.harvard.edu/abs/2013ApJ...778..178H} {778, 178}

\bibitem[\protect\citeauthoryear{{Inayoshi}, {Visbal}  \&
  {Kashiyama}}{{Inayoshi} et~al.}{2015}]{inayoshi15}
{Inayoshi} K.,  {Visbal} E.,   {Kashiyama} K.,  2015, \mn@doi [\mnras]
  {10.1093/mnras/stv1654}, \href
  {https://ui.adsabs.harvard.edu/abs/2015MNRAS.453.1692I} {453, 1692}

\bibitem[\protect\citeauthoryear{{Inayoshi}, {Haiman}  \&
  {Ostriker}}{{Inayoshi} et~al.}{2016}]{inayoshi16}
{Inayoshi} K.,  {Haiman} Z.,   {Ostriker} J.~P.,  2016, \mn@doi [\mnras]
  {10.1093/mnras/stw836}, \href
  {https://ui.adsabs.harvard.edu/abs/2016MNRAS.459.3738I} {459, 3738}

\bibitem[\protect\citeauthoryear{{Inayoshi}, {Li}  \& {Haiman}}{{Inayoshi}
  et~al.}{2018}]{inayoshi18}
{Inayoshi} K.,  {Li} M.,   {Haiman} Z.,  2018, \mn@doi [\mnras]
  {10.1093/mnras/sty1720}, \href
  {https://ui.adsabs.harvard.edu/abs/2018MNRAS.479.4017I} {479, 4017}

\bibitem[\protect\citeauthoryear{{Inayoshi}, {Visbal}  \& {Haiman}}{{Inayoshi}
  et~al.}{2020}]{inayoshi20}
{Inayoshi} K.,  {Visbal} E.,   {Haiman} Z.,  2020, \mn@doi [\araa]
  {10.1146/annurev-astro-120419-014455}, \href
  {https://ui.adsabs.harvard.edu/abs/2020ARA&A..58...27I} {58, 27}

\bibitem[\protect\citeauthoryear{{Katz}, {Sijacki}  \& {Haehnelt}}{{Katz}
  et~al.}{2015}]{katz15}
{Katz} H.,  {Sijacki} D.,   {Haehnelt} M.~G.,  2015, \mn@doi [\mnras]
  {10.1093/mnras/stv1048}, \href
  {https://ui.adsabs.harvard.edu/abs/2015MNRAS.451.2352K} {451, 2352}

\bibitem[\protect\citeauthoryear{{Kauffmann} \& {Haehnelt}}{{Kauffmann} \&
  {Haehnelt}}{2000}]{kauffmann00}
{Kauffmann} G.,  {Haehnelt} M.,  2000, \mn@doi [\mnras]
  {10.1046/j.1365-8711.2000.03077.x}, \href
  {https://ui.adsabs.harvard.edu/abs/2000MNRAS.311..576K} {311, 576}

\bibitem[\protect\citeauthoryear{{Kimura}, {Hosokawa}  \& {Sugimura}}{{Kimura}
  et~al.}{2020}]{kimura20}
{Kimura} K.,  {Hosokawa} T.,   {Sugimura} K.,  2020, arXiv e-prints, \href
  {https://ui.adsabs.harvard.edu/abs/2020arXiv201201452K} {p. arXiv:2012.01452}

\bibitem[\protect\citeauthoryear{{Koushiappas}, {Bullock}  \&
  {Dekel}}{{Koushiappas} et~al.}{2004}]{koushiappas04}
{Koushiappas} S.~M.,  {Bullock} J.~S.,   {Dekel} A.,  2004, \mn@doi [\mnras]
  {10.1111/j.1365-2966.2004.08190.x}, \href
  {http://adsabs.harvard.edu/abs/2004MNRAS.354..292K} {354, 292}

\bibitem[\protect\citeauthoryear{{Kroupa}, {Subr}, {Jerabkova}  \&
  {Wang}}{{Kroupa} et~al.}{2020}]{kroupa20}
{Kroupa} P.,  {Subr} L.,  {Jerabkova} T.,   {Wang} L.,  2020, \mn@doi [\mnras]
  {10.1093/mnras/staa2276}, \href
  {https://ui.adsabs.harvard.edu/abs/2020MNRAS.498.5652K} {498, 5652}

\bibitem[\protect\citeauthoryear{{Kulkarni}, {Visbal}  \& {Bryan}}{{Kulkarni}
  et~al.}{2020}]{kulkarni20}
{Kulkarni} M.,  {Visbal} E.,   {Bryan} G.~L.,  2020, arXiv e-prints, \href
  {https://ui.adsabs.harvard.edu/abs/2020arXiv201004169K} {p. arXiv:2010.04169}

\bibitem[\protect\citeauthoryear{{Lacey} \& {Cole}}{{Lacey} \&
  {Cole}}{1993}]{lacey93}
{Lacey} C.,  {Cole} S.,  1993, \mn@doi [\mnras] {10.1093/mnras/262.3.627},
  \href {https://ui.adsabs.harvard.edu/abs/1993MNRAS.262..627L} {262, 627}

\bibitem[\protect\citeauthoryear{{Latif}, {Schleicher}, {Schmidt}  \&
  {Niemeyer}}{{Latif} et~al.}{2013}]{latif13}
{Latif} M.~A.,  {Schleicher} D.~R.~G.,  {Schmidt} W.,   {Niemeyer} J.~C.,
  2013, \mn@doi [\mnras] {10.1093/mnras/stt1786}, \href
  {http://adsabs.harvard.edu/abs/2013MNRAS.436.2989L} {436, 2989}

\bibitem[\protect\citeauthoryear{{Latif}, {Bovino}, {Van Borm}, {Grassi},
  {Schleicher}  \& {Spaans}}{{Latif} et~al.}{2014}]{latif14}
{Latif} M.~A.,  {Bovino} S.,  {Van Borm} C.,  {Grassi} T.,  {Schleicher}
  D.~R.~G.,   {Spaans} M.,  2014, \mn@doi [\mnras] {10.1093/mnras/stu1230},
  \href {https://ui.adsabs.harvard.edu/abs/2014MNRAS.443.1979L} {443, 1979}

\bibitem[\protect\citeauthoryear{{Latif}, {Bovino}, {Grassi}, {Schleicher}  \&
  {Spaans}}{{Latif} et~al.}{2015}]{latif15}
{Latif} M.~A.,  {Bovino} S.,  {Grassi} T.,  {Schleicher} D.~R.~G.,   {Spaans}
  M.,  2015, \mn@doi [\mnras] {10.1093/mnras/stu2244}, \href
  {http://adsabs.harvard.edu/abs/2015MNRAS.446.3163L} {446, 3163}

\bibitem[\protect\citeauthoryear{{Latif}, {Lupi}, {Schleicher}, {D'Amico},
  {Panci}  \& {Bovino}}{{Latif} et~al.}{2019}]{latif19}
{Latif} M.~A.,  {Lupi} A.,  {Schleicher} D.~R.~G.,  {D'Amico} G.,  {Panci} P.,
   {Bovino} S.,  2019, \mn@doi [\mnras] {10.1093/mnras/stz608}, \href
  {https://ui.adsabs.harvard.edu/abs/2019MNRAS.485.3352L} {485, 3352}

\bibitem[\protect\citeauthoryear{{Lodato} \& {Natarajan}}{{Lodato} \&
  {Natarajan}}{2006}]{lodato06}
{Lodato} G.,  {Natarajan} P.,  2006, \mn@doi [\mnras]
  {10.1111/j.1365-2966.2006.10801.x}, \href
  {http://cdsads.u-strasbg.fr/abs/2006MNRAS.371.1813L} {371, 1813}

\bibitem[\protect\citeauthoryear{Lupi, Colpi, Devecchi, Galanti  \&
  Volonteri}{Lupi et~al.}{2014}]{lupi14}
Lupi A.,  Colpi M.,  Devecchi B.,  Galanti G.,   Volonteri M.,  2014, \mn@doi
  [Monthly Notices of the Royal Astronomical Society] {10.1093/mnras/stu1120},
  442, 3616

\bibitem[\protect\citeauthoryear{{Lupi}, {Haardt}, {Dotti}, {Fiacconi}, {Mayer}
   \& {Madau}}{{Lupi} et~al.}{2016}]{lupi16}
{Lupi} A.,  {Haardt} F.,  {Dotti} M.,  {Fiacconi} D.,  {Mayer} L.,   {Madau}
  P.,  2016, \mn@doi [\mnras] {10.1093/mnras/stv2877}, \href
  {https://ui.adsabs.harvard.edu/abs/2016MNRAS.456.2993L} {456, 2993}

\bibitem[\protect\citeauthoryear{{Lupi}, {Volonteri}, {Decarli}, {Bovino},
  {Silk}  \& {Bergeron}}{{Lupi} et~al.}{2019}]{lupi19b}
{Lupi} A.,  {Volonteri} M.,  {Decarli} R.,  {Bovino} S.,  {Silk} J.,
  {Bergeron} J.,  2019, \mn@doi [\mnras] {10.1093/mnras/stz1959}, \href
  {https://ui.adsabs.harvard.edu/abs/2019MNRAS.488.4004L} {488, 4004}

\bibitem[\protect\citeauthoryear{{Machacek}, {Bryan}  \& {Abel}}{{Machacek}
  et~al.}{2001}]{machacek01}
{Machacek} M.~E.,  {Bryan} G.~L.,   {Abel} T.,  2001, \mn@doi [\apj]
  {10.1086/319014}, \href {http://adsabs.harvard.edu/abs/2001ApJ...548..509M}
  {548, 509}

\bibitem[\protect\citeauthoryear{{Madau} \& {Rees}}{{Madau} \&
  {Rees}}{2001}]{madau01a}
{Madau} P.,  {Rees} M.~J.,  2001, \mn@doi [\apjl] {10.1086/319848}, \href
  {http://adsabs.harvard.edu/abs/2001ApJ...551L..27M} {551, L27}

\bibitem[\protect\citeauthoryear{{Madau}, {Ferrara}  \& {Rees}}{{Madau}
  et~al.}{2001}]{madau01b}
{Madau} P.,  {Ferrara} A.,   {Rees} M.~J.,  2001, \mn@doi [\apj]
  {10.1086/321474}, \href
  {https://ui.adsabs.harvard.edu/abs/2001ApJ...555...92M} {555, 92}

\bibitem[\protect\citeauthoryear{{Madau}, {Haardt}  \& {Dotti}}{{Madau}
  et~al.}{2014}]{madau14}
{Madau} P.,  {Haardt} F.,   {Dotti} M.,  2014, \mn@doi [\apjl]
  {10.1088/2041-8205/784/2/L38}, \href
  {http://adsabs.harvard.edu/abs/2014ApJ...784L..38M} {784, L38}

\bibitem[\protect\citeauthoryear{{Mayer}, {Kazantzidis}, {Escala}  \&
  {Callegari}}{{Mayer} et~al.}{2010}]{mayer10}
{Mayer} L.,  {Kazantzidis} S.,  {Escala} A.,   {Callegari} S.,  2010, \mn@doi
  [\nat] {10.1038/nature09294}, \href
  {http://cdsads.u-strasbg.fr/abs/2010Natur.466.1082M} {466, 1082}

\bibitem[\protect\citeauthoryear{{McKee} \& {Tan}}{{McKee} \&
  {Tan}}{2008}]{mckee08}
{McKee} C.~F.,  {Tan} J.~C.,  2008, \mn@doi [\apj] {10.1086/587434}, \href
  {http://adsabs.harvard.edu/abs/2008ApJ...681..771M} {681, 771}

\bibitem[\protect\citeauthoryear{{Miller} \& {Davies}}{{Miller} \&
  {Davies}}{2012}]{miller12}
{Miller} M.~C.,  {Davies} M.~B.,  2012, \mn@doi [\apj]
  {10.1088/0004-637X/755/1/81}, \href
  {http://cdsads.u-strasbg.fr/abs/2012ApJ...755...81M} {755, 81}

\bibitem[\protect\citeauthoryear{{Mortlock} et~al.,}{{Mortlock}
  et~al.}{2011}]{mortlock11}
{Mortlock} D.~J.,  et~al., 2011, \mn@doi [\nat] {10.1038/nature10159}, \href
  {http://adsabs.harvard.edu/abs/2011Natur.474..616M} {474, 616}

\bibitem[\protect\citeauthoryear{{Oh} \& {Haiman}}{{Oh} \&
  {Haiman}}{2002}]{oh02}
{Oh} S.~P.,  {Haiman} Z.,  2002, \mn@doi [\apj] {10.1086/339393}, \href
  {http://adsabs.harvard.edu/abs/2002ApJ...569..558O} {569, 558}

\bibitem[\protect\citeauthoryear{{Omukai}}{{Omukai}}{2001}]{omukai01a}
{Omukai} K.,  2001, \mn@doi [\apj] {10.1086/318296}, \href
  {https://ui.adsabs.harvard.edu/abs/2001ApJ...546..635O} {546, 635}

\bibitem[\protect\citeauthoryear{{Omukai} \& {Palla}}{{Omukai} \&
  {Palla}}{2001}]{omukai01}
{Omukai} K.,  {Palla} F.,  2001, \mn@doi [\apjl] {10.1086/324410}, \href
  {http://adsabs.harvard.edu/abs/2001ApJ...561L..55O} {561, L55}

\bibitem[\protect\citeauthoryear{{Omukai}, {Schneider}  \& {Haiman}}{{Omukai}
  et~al.}{2008}]{omukai08}
{Omukai} K.,  {Schneider} R.,   {Haiman} Z.,  2008, \mn@doi [\apj]
  {10.1086/591636}, \href {http://adsabs.harvard.edu/abs/2008ApJ...686..801O}
  {686, 801}

\bibitem[\protect\citeauthoryear{{Pacucci}, {Volonteri}  \&
  {Ferrara}}{{Pacucci} et~al.}{2015}]{pacucci15}
{Pacucci} F.,  {Volonteri} M.,   {Ferrara} A.,  2015, \mn@doi [\mnras]
  {10.1093/mnras/stv1465}, \href
  {https://ui.adsabs.harvard.edu/abs/2015MNRAS.452.1922P} {452, 1922}

\bibitem[\protect\citeauthoryear{{Peebles}}{{Peebles}}{1993}]{peebles93}
{Peebles} P.~J.~E.,  1993, {Principles of Physical Cosmology}.
Princeton University Press

\bibitem[\protect\citeauthoryear{{Pezzulli}, {Valiante}  \&
  {Schneider}}{{Pezzulli} et~al.}{2016}]{pezzulli16}
{Pezzulli} E.,  {Valiante} R.,   {Schneider} R.,  2016, \mn@doi [\mnras]
  {10.1093/mnras/stw505}, \href
  {https://ui.adsabs.harvard.edu/abs/2016MNRAS.458.3047P} {458, 3047}

\bibitem[\protect\citeauthoryear{{Planck Collaboration} et~al.,}{{Planck
  Collaboration} et~al.}{2016}]{planck16}
{Planck Collaboration} et~al., 2016, \mn@doi [A&A]
  {10.1051/0004-6361/201525830}, 594, A13

\bibitem[\protect\citeauthoryear{{Pollack}, {Spergel}  \&
  {Steinhardt}}{{Pollack} et~al.}{2015}]{pollack15}
{Pollack} J.,  {Spergel} D.~N.,   {Steinhardt} P.~J.,  2015, \mn@doi [\apj]
  {10.1088/0004-637X/804/2/131}, \href
  {https://ui.adsabs.harvard.edu/abs/2015ApJ...804..131P} {804, 131}

\bibitem[\protect\citeauthoryear{{Portegies Zwart} \& {McMillan}}{{Portegies
  Zwart} \& {McMillan}}{2002}]{portegies02}
{Portegies Zwart} S.~F.,  {McMillan} S.~L.~W.,  2002, \mn@doi [\apj]
  {10.1086/341798}, \href {http://cdsads.u-strasbg.fr/abs/2002ApJ...576..899P}
  {576, 899}

\bibitem[\protect\citeauthoryear{{Regan}, {Johansson}  \& {Wise}}{{Regan}
  et~al.}{2014}]{regan14}
{Regan} J.~A.,  {Johansson} P.~H.,   {Wise} J.~H.,  2014, \mn@doi [\apj]
  {10.1088/0004-637X/795/2/137}, \href
  {https://ui.adsabs.harvard.edu/abs/2014ApJ...795..137R} {795, 137}

\bibitem[\protect\citeauthoryear{{Regan}, {Visbal}, {Wise}, {Haiman},
  {Johansson}  \& {Bryan}}{{Regan} et~al.}{2017}]{regan17}
{Regan} J.~A.,  {Visbal} E.,  {Wise} J.~H.,  {Haiman} Z.,  {Johansson} P.~H.,
  {Bryan} G.~L.,  2017, \mn@doi [Nature Astronomy] {10.1038/s41550-017-0075},
  \href {https://ui.adsabs.harvard.edu/abs/2017NatAs...1E..75R} {1, 0075}

\bibitem[\protect\citeauthoryear{{Regan}, {Downes}, {Volonteri}, {Beckmann},
  {Lupi}, {Trebitsch}  \& {Dubois}}{{Regan} et~al.}{2019}]{regan19}
{Regan} J.~A.,  {Downes} T.~P.,  {Volonteri} M.,  {Beckmann} R.,  {Lupi} A.,
  {Trebitsch} M.,   {Dubois} Y.,  2019, \mn@doi [\mnras]
  {10.1093/mnras/stz1045}, \href
  {https://ui.adsabs.harvard.edu/abs/2019MNRAS.tmp..999R} {}

\bibitem[\protect\citeauthoryear{{Regan}, {Haiman}, {Wise}, {O'Shea}  \&
  {Norman}}{{Regan} et~al.}{2020a}]{regan20}
{Regan} J.~A.,  {Haiman} Z.,  {Wise} J.~H.,  {O'Shea} B.~W.,   {Norman} M.~L.,
  2020a, \mn@doi [The Open Journal of Astrophysics]
  {10.21105/astro.2006.14625}, \href
  {https://ui.adsabs.harvard.edu/abs/2020OJAp....3E...9R} {3, E9}

\bibitem[\protect\citeauthoryear{{Regan}, {Wise}, {Woods}, {Downes}, {O'Shea}
  \& {Norman}}{{Regan} et~al.}{2020b}]{regan20b}
{Regan} J.~A.,  {Wise} J.~H.,  {Woods} T.~E.,  {Downes} T.~P.,  {O'Shea} B.~W.,
    {Norman} M.~L.,  2020b, \mn@doi [The Open Journal of Astrophysics]
  {10.21105/astro.2008.08090}, \href
  {https://ui.adsabs.harvard.edu/abs/2020OJAp....3E..15R} {3, 15}

\bibitem[\protect\citeauthoryear{{Reinoso}, {Schleicher}, {Fellhauer},
  {Klessen}  \& {Boekholt}}{{Reinoso} et~al.}{2018}]{reinoso18}
{Reinoso} B.,  {Schleicher} D.~R.~G.,  {Fellhauer} M.,  {Klessen} R.~S.,
  {Boekholt} T.~C.~N.,  2018, \mn@doi [\aap] {10.1051/0004-6361/201732224},
  \href {https://ui.adsabs.harvard.edu/abs/2018A&A...614A..14R} {614, A14}

\bibitem[\protect\citeauthoryear{{Sakurai}, {Inayoshi}  \& {Haiman}}{{Sakurai}
  et~al.}{2016}]{sakurai16}
{Sakurai} Y.,  {Inayoshi} K.,   {Haiman} Z.,  2016, \mn@doi [\mnras]
  {10.1093/mnras/stw1652}, \href
  {https://ui.adsabs.harvard.edu/abs/2016MNRAS.461.4496S} {461, 4496}

\bibitem[\protect\citeauthoryear{{Sakurai}, {Haiman}  \& {Inayoshi}}{{Sakurai}
  et~al.}{2020}]{sakurai20}
{Sakurai} Y.,  {Haiman} Z.,   {Inayoshi} K.,  2020, \mn@doi [\mnras]
  {10.1093/mnras/staa3227}, \href
  {https://ui.adsabs.harvard.edu/abs/2020MNRAS.499.5960S} {499, 5960}

\bibitem[\protect\citeauthoryear{{Schaerer}}{{Schaerer}}{2003}]{schaerer03}
{Schaerer} D.,  2003, \mn@doi [\aap] {10.1051/0004-6361:20021525}, \href
  {https://ui.adsabs.harvard.edu/abs/2003A&A...397..527S} {397, 527}

\bibitem[\protect\citeauthoryear{{Schauer} et~al.,}{{Schauer}
  et~al.}{2017}]{schauer17}
{Schauer} A. T.~P.,  et~al., 2017, \mn@doi [\mnras] {10.1093/mnras/stx264},
  \href {https://ui.adsabs.harvard.edu/abs/2017MNRAS.467.2288S} {467, 2288}

\bibitem[\protect\citeauthoryear{Schauer, Glover, Klessen  \& Ceverino}{Schauer
  et~al.}{2019}]{schauer19}
Schauer A. T.~P.,  Glover S. C.~O.,  Klessen R.~S.,   Ceverino D.,  2019,
  \mn@doi [Monthly Notices of the Royal Astronomical Society]
  {10.1093/mnras/stz013}, 484, 3510

\bibitem[\protect\citeauthoryear{{Schlaufman}, {Thompson}  \&
  {Casey}}{{Schlaufman} et~al.}{2018}]{schlaufman2018}
{Schlaufman} K.~C.,  {Thompson} I.~B.,   {Casey} A.~R.,  2018, \mn@doi [\apj]
  {10.3847/1538-4357/aadd97}, \href
  {https://ui.adsabs.harvard.edu/abs/2018ApJ...867...98S} {867, 98}

\bibitem[\protect\citeauthoryear{{Shang}, {Bryan}  \& {Haiman}}{{Shang}
  et~al.}{2010}]{shang10}
{Shang} C.,  {Bryan} G.~L.,   {Haiman} Z.,  2010, \mn@doi [\mnras]
  {10.1111/j.1365-2966.2009.15960.x}, \href
  {https://ui.adsabs.harvard.edu/abs/2010MNRAS.402.1249S} {402, 1249}

\bibitem[\protect\citeauthoryear{{Shankar}, {Weinberg}  \&
  {Miralda-Escud{\'e}}}{{Shankar} et~al.}{2009}]{shankar09}
{Shankar} F.,  {Weinberg} D.~H.,   {Miralda-Escud{\'e}} J.,  2009, \mn@doi
  [\apj] {10.1088/0004-637X/690/1/20}, \href
  {https://ui.adsabs.harvard.edu/abs/2009ApJ...690...20S} {690, 20}

\bibitem[\protect\citeauthoryear{{Shen}, {Hopkins}, {Faucher-Gigu{\`e}re},
  {Alexander}, {Richards}, {Ross}  \& {Hickox}}{{Shen} et~al.}{2020}]{shen20}
{Shen} X.,  {Hopkins} P.~F.,  {Faucher-Gigu{\`e}re} C.-A.,  {Alexander} D.~M.,
  {Richards} G.~T.,  {Ross} N.~P.,   {Hickox} R.~C.,  2020, \mn@doi [\mnras]
  {10.1093/mnras/staa1381}, \href
  {https://ui.adsabs.harvard.edu/abs/2020MNRAS.495.3252S} {495, 3252}

\bibitem[\protect\citeauthoryear{{Small} \& {Blandford}}{{Small} \&
  {Blandford}}{1992}]{small92}
{Small} T.~A.,  {Blandford} R.~D.,  1992, \mn@doi [\mnras]
  {10.1093/mnras/259.4.725}, \href
  {https://ui.adsabs.harvard.edu/abs/1992MNRAS.259..725S} {259, 725}

\bibitem[\protect\citeauthoryear{{Soltan}}{{Soltan}}{1982}]{soltan82}
{Soltan} A.,  1982, \mnras, \href
  {http://adsabs.harvard.edu/abs/1982MNRAS.200..115S} {200, 115}

\bibitem[\protect\citeauthoryear{{Springel}}{{Springel}}{2005}]{springel05}
{Springel} V.,  2005, \mn@doi [\mnras] {10.1111/j.1365-2966.2005.09655.x},
  \href {http://adsabs.harvard.edu/abs/2005MNRAS.364.1105S} {364, 1105}

\bibitem[\protect\citeauthoryear{{Stacy}, {Greif}  \& {Bromm}}{{Stacy}
  et~al.}{2012}]{stacy12}
{Stacy} A.,  {Greif} T.~H.,   {Bromm} V.,  2012, \mn@doi [\mnras]
  {10.1111/j.1365-2966.2012.20605.x}, \href
  {http://adsabs.harvard.edu/abs/2012MNRAS.422..290S} {422, 290}

\bibitem[\protect\citeauthoryear{{Tagawa}, {Umemura}, {Gouda}, {Yano}  \&
  {Yamai}}{{Tagawa} et~al.}{2015}]{tagawa15}
{Tagawa} H.,  {Umemura} M.,  {Gouda} N.,  {Yano} T.,   {Yamai} Y.,  2015,
  \mn@doi [\mnras] {10.1093/mnras/stv1099}, \href
  {https://ui.adsabs.harvard.edu/abs/2015MNRAS.451.2174T} {451, 2174}

\bibitem[\protect\citeauthoryear{{Tagawa}, {Haiman}  \& {Kocsis}}{{Tagawa}
  et~al.}{2020}]{tagawa20}
{Tagawa} H.,  {Haiman} Z.,   {Kocsis} B.,  2020, \mn@doi [\apj]
  {10.3847/1538-4357/ab7922}, \href
  {https://ui.adsabs.harvard.edu/abs/2020ApJ...892...36T} {892, 36}

\bibitem[\protect\citeauthoryear{{Tanaka} \& {Haiman}}{{Tanaka} \&
  {Haiman}}{2009}]{tanaka09}
{Tanaka} T.,  {Haiman} Z.,  2009, \mn@doi [\apj]
  {10.1088/0004-637X/696/2/1798}, \href
  {http://adsabs.harvard.edu/abs/2009ApJ...696.1798T} {696, 1798}

\bibitem[\protect\citeauthoryear{{Tegmark}, {Silk}, {Rees}, {Blanchard}, {Abel}
   \& {Palla}}{{Tegmark} et~al.}{1997}]{tegmark97}
{Tegmark} M.,  {Silk} J.,  {Rees} M.~J.,  {Blanchard} A.,  {Abel} T.,   {Palla}
  F.,  1997, \mn@doi [\apj] {10.1086/303434}, \href
  {http://adsabs.harvard.edu/abs/1997ApJ...474....1T} {474, 1}

\bibitem[\protect\citeauthoryear{{Trenti}, {Stiavelli}  \& {Michael
  Shull}}{{Trenti} et~al.}{2009}]{trenti09}
{Trenti} M.,  {Stiavelli} M.,   {Michael Shull} J.,  2009, \mn@doi [\apj]
  {10.1088/0004-637X/700/2/1672}, \href
  {http://adsabs.harvard.edu/abs/2009ApJ...700.1672T} {700, 1672}

\bibitem[\protect\citeauthoryear{{Valiante}, {Schneider}, {Volonteri}  \&
  {Omukai}}{{Valiante} et~al.}{2016}]{valiante16}
{Valiante} R.,  {Schneider} R.,  {Volonteri} M.,   {Omukai} K.,  2016, \mn@doi
  [\mnras] {10.1093/mnras/stw225}, \href
  {https://ui.adsabs.harvard.edu/abs/2016MNRAS.457.3356V} {457, 3356}

\bibitem[\protect\citeauthoryear{{Visbal}, {Haiman}  \& {Bryan}}{{Visbal}
  et~al.}{2014a}]{visbal14a}
{Visbal} E.,  {Haiman} Z.,   {Bryan} G.~L.,  2014a, \mn@doi [\mnras]
  {10.1093/mnrasl/slu063}, \href
  {https://ui.adsabs.harvard.edu/abs/2014MNRAS.442L.100V} {442, L100}

\bibitem[\protect\citeauthoryear{{Visbal}, {Haiman}  \& {Bryan}}{{Visbal}
  et~al.}{2014b}]{visbal14b}
{Visbal} E.,  {Haiman} Z.,   {Bryan} G.~L.,  2014b, \mn@doi [\mnras]
  {10.1093/mnras/stu1794}, \href
  {https://ui.adsabs.harvard.edu/abs/2014MNRAS.445.1056V} {445, 1056}

\bibitem[\protect\citeauthoryear{{Volonteri} \& {Rees}}{{Volonteri} \&
  {Rees}}{2005}]{volonteri05}
{Volonteri} M.,  {Rees} M.~J.,  2005, \mn@doi [\apj] {10.1086/466521}, \href
  {https://ui.adsabs.harvard.edu/abs/2005ApJ...633..624V} {633, 624}

\bibitem[\protect\citeauthoryear{{Volonteri}, {Haardt}  \& {Madau}}{{Volonteri}
  et~al.}{2003}]{volonteri03}
{Volonteri} M.,  {Haardt} F.,   {Madau} P.,  2003, \mn@doi [\apj]
  {10.1086/344675}, \href {http://adsabs.harvard.edu/abs/2003ApJ...582..559V}
  {582, 559}

\bibitem[\protect\citeauthoryear{{Volonteri}, {Silk}  \& {Dubus}}{{Volonteri}
  et~al.}{2015}]{volonteri15}
{Volonteri} M.,  {Silk} J.,   {Dubus} G.,  2015, \mn@doi [\apj]
  {10.1088/0004-637X/804/2/148}, \href
  {http://adsabs.harvard.edu/abs/2015ApJ...804..148V} {804, 148}

\bibitem[\protect\citeauthoryear{{Watson}, {Iliev}, {D'Aloisio}, {Knebe},
  {Shapiro}  \& {Yepes}}{{Watson} et~al.}{2013}]{watson13}
{Watson} W.~A.,  {Iliev} I.~T.,  {D'Aloisio} A.,  {Knebe} A.,  {Shapiro} P.~R.,
    {Yepes} G.,  2013, \mn@doi [\mnras] {10.1093/mnras/stt791}, \href
  {https://ui.adsabs.harvard.edu/abs/2013MNRAS.433.1230W} {433, 1230}

\bibitem[\protect\citeauthoryear{{Weaver}, {McCray}, {Castor}, {Shapiro}  \&
  {Moore}}{{Weaver} et~al.}{1977}]{weaver77}
{Weaver} R.,  {McCray} R.,  {Castor} J.,  {Shapiro} P.,   {Moore} R.,  1977,
  \mn@doi [\apj] {10.1086/155692}, \href
  {https://ui.adsabs.harvard.edu/abs/1977ApJ...218..377W} {218, 377}

\bibitem[\protect\citeauthoryear{{Wise}, {Regan}, {O'Shea}, {Norman}, {Downes}
  \& {Xu}}{{Wise} et~al.}{2019}]{wise19}
{Wise} J.~H.,  {Regan} J.~A.,  {O'Shea} B.~W.,  {Norman} M.~L.,  {Downes}
  T.~P.,   {Xu} H.,  2019, \mn@doi [\nat] {10.1038/s41586-019-0873-4}, \href
  {https://ui.adsabs.harvard.edu/abs/2019Natur.566...85W} {566, 85}

\bibitem[\protect\citeauthoryear{{Wolcott-Green} \& {Haiman}}{{Wolcott-Green}
  \& {Haiman}}{2019}]{wolcott-green19}
{Wolcott-Green} J.,  {Haiman} Z.,  2019, \mn@doi [\mnras]
  {10.1093/mnras/sty3280}, \href
  {https://ui.adsabs.harvard.edu/abs/2019MNRAS.484.2467W} {484, 2467}

\bibitem[\protect\citeauthoryear{{Wolcott-Green}, {Haiman}  \&
  {Bryan}}{{Wolcott-Green} et~al.}{2011}]{wolcott-green11}
{Wolcott-Green} J.,  {Haiman} Z.,   {Bryan} G.~L.,  2011, \mn@doi [\mnras]
  {10.1111/j.1365-2966.2011.19538.x}, \href
  {https://ui.adsabs.harvard.edu/abs/2011MNRAS.418..838W} {418, 838}

\bibitem[\protect\citeauthoryear{{Wollenberg}, {Glover}, {Clark}  \&
  {Klessen}}{{Wollenberg} et~al.}{2020}]{wollenberg20}
{Wollenberg} K. M.~J.,  {Glover} S. C.~O.,  {Clark} P.~C.,   {Klessen} R.~S.,
  2020, \mn@doi [\mnras] {10.1093/mnras/staa289}, \href
  {https://ui.adsabs.harvard.edu/abs/2020MNRAS.494.1871W} {494, 1871}

\bibitem[\protect\citeauthoryear{{Woods}, {Heger}, {Whalen}, {Haemmerl{\'e}}
  \& {Klessen}}{{Woods} et~al.}{2017}]{woods17}
{Woods} T.~E.,  {Heger} A.,  {Whalen} D.~J.,  {Haemmerl{\'e}} L.,   {Klessen}
  R.~S.,  2017, \mn@doi [\apjl] {10.3847/2041-8213/aa7412}, \href
  {https://ui.adsabs.harvard.edu/abs/2017ApJ...842L...6W} {842, L6}

\bibitem[\protect\citeauthoryear{{Yoshida}, {Abel}, {Hernquist}  \&
  {Sugiyama}}{{Yoshida} et~al.}{2003}]{yoshida03}
{Yoshida} N.,  {Abel} T.,  {Hernquist} L.,   {Sugiyama} N.,  2003, \mn@doi
  [\apj] {10.1086/375810}, \href
  {https://ui.adsabs.harvard.edu/abs/2003ApJ...592..645Y} {592, 645}

\makeatother
\end{thebibliography}







\bsp	
\label{lastpage}
\end{document}